\documentclass[a4paper,10pt]{article}
\usepackage{jheppub} % for details on the use of the package, please
\usepackage{bm}
\usepackage{amsmath,amssymb,bm}
\usepackage{graphicx}
\usepackage[utf8]{inputenc}
\usepackage{verbatim}

\numberwithin{equation}{section}

\usepackage{cancel}

\newcommand{\be}{\begin{equation}}
\newcommand{\ee}{\end{equation}}
\newcommand{\bea}{\begin{eqnarray}}
\newcommand{\eea}{\end{eqnarray}}

\newcommand{\eq}[1]{Eq.~(\ref{eq:#1})}
\newcommand{\sect}[1]{Sec.~\ref{sec:#1}}
\newcommand{\appen}[1]{Appendix~\ref{sec:#1}}

\newcommand{\del}{\partial}

\newcommand{\Tc}{T_c}

\newcommand{\bra}{\langle}
\newcommand{\ket}{\rangle}

\newcommand{\eg}{{\it e.g.}}

\newcommand{\bh}{black hole\ }

\newcommand{\SC}{superconductor}
\newcommand{\SCs}{superconductors}
\newcommand{\HSC}{holographic superconductor}
\newcommand{\HSCs}{holographic superconductors}

\newcommand{\margin}[1]{\marginpar{\tiny #1}}

\bmdefine{\bmq}{{\bm{q}}}
\bmdefine{\bmk}{{\bm{k}}}
\bmdefine{\bmx}{{\bm{x}}}
\bmdefine{\bmy}{{\bm{y}}}
\bmdefine{\bmr}{{\bm{r}}}
\bmdefine{\bmnabla}{{\bm{\nabla}}}
\bmdefine{\bmA}{ \bm{A} }
\bmdefine{\bmD}{ \bm{D} }

\bmdefine{\bmPhi}{ \bm{\Phi} }
\bmdefine{\bmPsi}{ \bm{\Psi} }
\bmdefine{\bmcalO}{ \bm{\mathcal{O}} }
\bmdefine{\bmrho}{ \bm{\rho} }

\newcommand{\vecx}{\vec{x}}

\newcommand{\epsmu}{\epsilon_{\mu}}

\newcommand{\calP}{{\cal P}}

\bmdefine{\bmg}{{\bm{g}}}
\bmdefine{\bmR}{{\bm{R}}}

\newcommand{\mfA}{\mathfrak{A}}

\newcommand{\half}{\frac{1}{2}}

\newcommand{\calA}{{\cal A}}

\newcommand{\calF}{{\cal F}}

\newcommand{\calJ}{{\cal J}}

\newcommand{\mass}{\text{M}}

\newcommand{\mfa}{\mathfrak{a}}
\newcommand{\epse}{\varepsilon_e}
\newcommand{\mfat}{\mfA_{t}}
\newcommand{\Cone}{\delta\psi}

\newcommand{\eps}{\epsilon}

\newcommand{\SFs}{superfluids}
\newcommand{\HSF}{holographic superfluid}
\newcommand{\HSFs}{holographic superfluids}

\newcommand{\At}{\calA_t}
\newcommand{\dpsi}{\delta\psi}

\newcommand{\Jc}{J^*}

\newcommand{\calQ}{{\cal Q}}
\newcommand{\HSFHSC}{holographic superfluid/superconductor}
\newcommand{\HSFHSCs}{holographic superfluids/superconductors}
\newcommand{\SFSC}{superfluid/superconductor}
\title{The dual Ginzburg-Landau theory for a holographic superfluid/superconductor: Critical dynamics}%

\author[a,1]{Makoto Natsuume%
\note{Also at Graduate Institute for Advanced Studies, 
SOKENDAI, 1-1 Oho, 
Tsukuba, Ibaraki, 305-0801, Japan;
Department of Mechanical Engineering, Mie University, Tsu, 514-8507, Japan.}
}
\affiliation[a]{KEK Theory Center, \\
Institute of Particle and Nuclear Studies, \\
High Energy Accelerator Research Organization,\\
Tsukuba, Ibaraki, 305-0801, Japan}
\emailAdd{makoto.natsuume@kek.jp}
\arxivnumber{2605.20415}
\preprint{KEK-TH-2784} 
\abstract{
Holographic superfluids/superconductors are one of the most studied systems in the AdS/CFT duality. In the low-energy, in the long-wavelength limit, they should be described by a Ginzburg-Landau theory. For critical dynamics, one expects that they belong to ``model F" universality class.
We consider a bulk 5-dimensional holographic superfluid/superconductor in the probe limit. For the holographic superconductor, we impose the boundary Maxwell equation to make the boundary Maxwell field dynamical. We identify the dual model F equations where numerical coefficients are obtained exactly. 
}

\keywords{Holography and condensed matter physics (AdS/CMT), AdS-CFT Correspondence, Black Holes}

\begin{document}

%\begin{flushright}
%        \today
%        {\it Preliminary version}
%\end{flushright}

\maketitle
\flushbottom

%
%\pacs{11.25.Tq, 64.60Ht, 25.75.-q}%, 74.20.-z} %?
% KEK-TH-XXXX

%\tableofcontents

%\newpage
%%%%%%%%%
\section{Introduction and Summary}%\label{sec:}
%%%%%%%%%

Holographic \SCs\ and \HSFs\ are one of the most studied systems in the AdS/CFT duality or the holographic duality \cite{Maldacena:1997re,Witten:1998qj,Witten:1998zw,Gubser:1998bc}.%
\footnote{For a textbook, see, \eg, Refs.~\cite{CasalderreySolana:2011us,Natsuume:2014sfa,Ammon:2015wua,Zaanen:2015oix,Hartnoll:2016apf,Baggioli:2019rrs}} 
They describe \SCs\ and \SFs\ \cite{Gubser:2008px,Hartnoll:2008vx,Hartnoll:2008kx}. On the other hand, from the macroscopic point of view, such a system should be described by the Ginzburg-Landau (GL) theory near the critical point. 

However, the exact form of the dual GL theory is little known. This is because a \HSF/\SC\ is typically an Einstein-Maxwell-complex scalar system. Such a system is hard to solve in general, and one needs a numerical computation. Even if one uses the probe limit where the backreation of matter fields onto the geometry is ignored, one has to solve a Maxwell-complex scalar system in an AdS \bh background, and one still needs a numerical computation. 

The exact form of the dual GL theory was identified only recently. In the bulk 5-dimensions, one can construct analytic solutions near the critical point with a particular bulk scalar mass \cite{Herzog:2010vz}. Using the solution and its generalizations, we identify the dual GL theory for the following systems:
\begin{itemize}
\item
A ``minimal" \HSC\ \cite{Natsuume:2018yrg,Natsuume:2022kic}.
\item
A class of ``nonminimal" \HSCs\ \cite{first}.
\item
A class of ``nonminimal" \HSCs\ at finite 't~Hooft coupling (Gauss-Bonnet \HSC) \cite{2nd,2nd2}.
\item
A class of holographic Lifshitz \SCs\ \cite{Natsuume:2018yrg}
\end{itemize}
Here, all numerical coefficients of the dual GL theories are obtained exactly. 

Traditionally, most literature call both \HSFs/\SCs\ as ``\HSCs" collectively, but we make the clear distinction between these two systems.
The distinction lies in the difference whether the boundary Maxwell field is dynamical or not.
%\margin{E OK}
\begin{enumerate}
\item
If the Maxwell field is dynamical, the system is a \HSC. 
\item
If the Maxwell field is added just as an external source, the system is  a \HSF. 
\end{enumerate}
Most literature actually study \HSFs.

From the holographic point of view, one can make the boundary Maxwell field dynamical by changing the boundary condition at the AdS boundary \cite{Natsuume:2022kic}.  
In this sense, we study both \HSC\ and \HSF\ in this paper.
As the result of the dynamical Maxwell equation, the hallmark of the \SC, the Meissner effect arises. 

However, the dual GL theories mentioned above are the stationary case.  It is natural to ask the time-dependence of a \HSFHSC. A critical dynamics is classified by a dynamic universality class (\sect{class}). 
The time-dependence of a \HSFHSC\ is even harder to solve than the stationary case. However, it is commonly believed that critical dynamics of a \HSF\ is described by ``model F" equations. The model F describes a superfluid. 
% v2
There is some recent progress, most notably Refs.~\cite{Flory:2022uzp,Donos:2022qao,Donos:2025jxb}. 
%\margin{E OK}
%v2
Ref.~\cite{Flory:2022uzp} relies on a numerical computation to compare model F equation and a holographic computation. 
Refs.~\cite{Donos:2022qao,Donos:2025jxb} show a formal proof of the equivalence of a holographic computation and model F equation. 
%These studies rely on numerical computations to compare model F equations and holographic computations. 
But it is desirable to obtain an exact explicit example.
The purpose of this paper is to identify the dual model F equations for a minimal \HSFHSC.%
\footnote{While this paper is in preparation, there appears a preprint which studies the same system as ours \cite{Bu:2026mxr}. See \appen{comparison} for the comparison with our results. }

%%------------------
\subsection{Summary}%\label{sec:}
%%------------------

%A \HSC\ is parameterized by a dimensionless parameter $\mu/T$, where $\mu$ is the chemical potential and $T$ is the temperature. We fix $T$ and vary $\mu$. Our results are summarized by the following TDGL equation in the unit $(\pi T)=1$:

%In this section, we review the dual GL theory for a minimal \HSC\ based on Refs.~XX.
\begin{itemize}
\item
We consider a bulk 5-dimensional minimal \HSFHSC\ which corresponds to a 4-dimensional \SFSC.  As in our previous analysis \cite{Natsuume:2018yrg,Natsuume:2022kic,first,2nd}, we consider the bulk scalar mass with $m^2=-4/L^2$.
%which saturates the Breitenlohner-Freedman (BF) bound \cite{Breitenlohner:1982bm}. 
In this case, a simple analytic solution is available at the critical point \cite{Herzog:2010vz}.
\item
A \HSFHSC\ has 2 dimensionful control parameters $\mu,T$, where $\mu$ is the chemical potential and $T$ is the temperature. Then, the system is parameterized by a dimensionless parameter $\mu/T$. We fix $T$ and vary $\mu$. 
\item
Also, we set the AdS radius $L=1$ and the horizon radius $r_0=1$ (or the Hawking temperature $\pi T=1$). 
\end{itemize}
% v12
In the stationary case, the dual GL theory is given by \cite{Natsuume:2018yrg,Natsuume:2022kic}
\begin{subequations}
\label{eq:GL4}
\begin{align}
f&=c_K |D_i\psi|^2 - a|\psi|^2+\frac{b}{2}|\psi|^4+\cdots+\frac{1}{4\mu_m}\calF_{ij}^2-(\psi J^*+\psi^* J)~, \\
%f &= \frac{1}{4} |D_i\psi|^2 - \half\epsmu|\psi|^2+\frac{1}{96}|\psi|^4+\cdots+\frac{1}{4\mu_m}\calF_{ij}^2-(\psi J^*+\psi^* J)~, \\
D_i &=\del_i-i\calA_i~, \\
a &=\frac{1}{2}\epsmu+\cdots~, \quad
b = \frac{1}{48} +\cdots~, \quad
c_K = \frac{1}{4} +\cdots~, \\
\frac{1}{\mu_m} &=\frac{1}{e^2}-\ln(\pi T)~.
%\label{eq:}
\end{align}
\end{subequations}
% v2
Here, $f$ is the free energy density, $x^i=(x,y,z)$ and
\begin{itemize}
\item
$\epsmu:=\mu-\mu_c$ is the deviation of the chemical potential from the critical point $\mu_c=2$. 
\item
The Maxwell action is added for the \HSC.
$e$ is the $U(1)$ coupling, and 
$\mu_m$ is the magnetic permeability due to the normal current or the medium effect \cite{Natsuume:2022kic}. In the standard GL theory, $\mu_m=e^2$.
\item
The $T$-dependence is shown explicitly only for the $\ln (\pi T)$ term.
\item
The coefficient $c_K$ is actually redundant because one can always absorb it by a $\psi$ scaling. Thus, there are 2 independent parameters $a,b$. But it is useful to keep it to compare with holographic results. This is because the dual GL theory typically does not have the canonical normalization if one uses the standard AdS/CFT dictionary. 
\end{itemize}
This free energy should be regarded as leading terms in the effective theory expansion. There should be the $O(|\psi|^6)$ term and higher, and numerical coefficients $a,b,c_K$ are the leading ones. 

In the time-dependent case, the system is dissipative so that the system cannot be written only by a free energy unless one uses the Martin-Siggia-Rose (MSR) formalism 
% v2
(see, \eg, Ref.~\cite{cardy}). 
It is traditional to define the system by a set of equations of motion.
% v2
Our model F equations are given by 
\begin{subequations}
%\label{eq:F}
\begin{align}
D_t\psi &= - \tau_0^{-1}\frac{\delta F}{\delta \psi^*}  -ig_0 \psi\frac{\delta F}{\delta \rho}~, \\
\del_t \rho &= \lambda \del_i^2 \left( \frac{\delta F}{\delta \rho} \right) - \lambda \del_iE^i  +2g_0 \Im \left( \psi^*\frac{\delta F}{\delta \psi^*} \right)~,
%\label{eq:}
%
\end{align}
\end{subequations}
as well as the complex conjugate equation for $\psi^*$. 
Here, $D_\mu=\del_\mu-ig_0 \calA_\mu$ and $E^i$ is the electric field $E^i=\calF_{it} =\del_i \calA_t- \del_t \calA_i$. $F$ is the GL free energy $F=\int d^3x f$ given by
\begin{align}
f &= c_K|D_i\psi|^2 - a |\psi|^2 + \frac{b'}{2}|\psi|^4+
\frac{1}{2C}\rho^2 +\gamma \rho |\psi|^2
%-\rho \At 
-(\psi J^*+\psi^* J)~. 
\label{eq:F_free}
\end{align}
Again, these equations should be regarded as leading terms in the effective theory expansion. 

A few remarks are in order:
\begin{itemize}
\item
We identify $\rho$ as the charge density and $\calA_t$ as the gauge potential. $\calA_i$ is the vector potential.
\item
Because we identify model F as the one with the Maxwell field, we slightly modify model F equations so that the system includes both $\calA_t$ and $\calA_i$ (\appen{conventional}). We also write down equations so that the $U(1)$ gauge symmetry is manifest (\sect{F}).
\item
The model has 5 additional parameters $\tau_0, D,C,g_0,\gamma$ in addition to the stationary case parameters $c_K,a,b$.
$\tau_0$ is the diffusion constant for $\psi$, 
% v2
$C:=\del\rho/\del\mu$ is the charge susceptibility, $\lambda$ is the conductivity, and $D:=\lambda/C$ is the diffusion constant for $\rho$. Comparing with our holographic results, we obtain%
\footnote{
The readers may notice that the coefficient of the $|\psi|^4$ term, $b'$, is different from $b$ in the stationary case \eqref{eq:GL4}.
But in the stationary case, model F equations reduce to the free energy \eqref{eq:GL4} (\sect{F_to_GL}). 
To avoid the confusion, we use the convention $b'$ here.
}
\begin{align}
\tau_0=\frac{1-3i}{4},
\quad
D=\half, 
\quad
C=2, 
\quad
g_0=1,
\quad
\gamma=-\frac{1}{4},
\quad
a= \half \epsmu,
\quad
b'=\frac{7}{48}~.
%\label{eq:}
%
\end{align}
%\margin{$e=g_0$?}

\item
Solving these equations perturbatively gives 3 hydrodynamic poles that correspond to the Maxwell diffusion mode and the second sound modes. 
\item
Model F equations may look complicated, but if one considers the high-temperature phase and if one focuses on the order parameter dynamics only, the system can be written by a time-dependent GL (TDGL) equation (\sect{high}).

\item 
In order to obtain model F parameters, one usually solves the full time-dependent problems in the low-temperature phase. But it is not really necessary. As we argue below, these parameters can be determined by solving much simpler problems (high-temperature phase and the low-temperature background). 
%\margin{TBC}
This is special to model F near the critical point. This would not be the case away from the critical point in general. 
\item
But, of course, \textit{it is a different issue if critical dynamics of \HSFHSC\ really follows model F predictions in the low-temperature phase}, so we explicitly solve the low-temperature phase and confirm this.

\item
Thus, our suggestion is that one first solves these simpler problems and determine model F parameters and solve the full time-dependent problem in the low-temperature phase. Then, compare model F and holographic results. 
% Any deviation from model F should be explained by subleading terms.

\end{itemize}
For the \HSC, the boundary Maxwell field must be dynamical:
\begin{itemize}

\item
Thus, for \HSCs, we add the boundary Maxwell equation as the boundary condition at the AdS boundary:
\begin{align}
\del_\nu \calF^{\mu\nu} =e^2 \bra \calJ^\mu \ket~.
%\label{eq:}
%
\end{align}
\item
Similarly, in the dual GL theory, we add the following boundary Maxwell action to the free energy \eqref{eq:F_free}:
\begin{align}
\delta f = \frac{1}{4e^2}\calF^{\mu\nu}\calF_{\mu\nu}~.
%\delta f = \frac{\epse(k)}{2}\calF^{ti}\calF_{ti} + \frac{1}{4\mu_m(k)}\calF^{ij}\calF_{ij}~.
%\label{eq:}
%
\end{align}
%\margin{F includes normal current, vector does not}
However, model F includes the normal current, so  the medium effect changes $e^2$ from the vacuum value to the electric permittivity $\epse(k)$ and the magnetic permeability $\mu_m(k)$.
%$\epse(k)$ is the electric permittivity and $\mu_m(k)$ is the magnetic permeability due to the normal current.  
In general, they are $\omega, q$-dependent, and their explicit form are given in \sect{epse}.

\item 
Solving model F equations and the boundary Maxwell equation now gives 2 hydrodynamic poles.
%\item
%However, the left-hand-side of the $t$-component of the Maxwell equation does not contribute at the leading order, and the equation reduces to the Neumann boundary condition $\bra\calJ^t\ket=0$.

%\item
%This has an interesting implication. For the \HSF, we obtain
%\begin{align}
%
%\delta\calJ^t = 2q^2 \frac{\calQ}{\calP} \mfA_t+\cdots~.
%\label{eq:}
%
%\end{align}
%where the scalar mode has a common denominator $\calP$ that gives 3 poles: the Maxwell diffusion mode and second sound modes. As the result of the Neumann boundary condition,
%\begin{align}
%
%\mfA_t\propto -\frac{\calP}{\calQ} \calJ^t_\text{ext} +\cdots~.
%\label{eq:}
%
%\end{align}
%where the scalar mode has a common denominator $\calQ$ that gives 2 dissipative poles. 

%\item
%Thus, the result of the \HSF\ implicitly contains the result of the \HSC. One would say that the denominator $\calP$ gives poles as the \HSF\ whereas the numerator $\calQ$ gives poles as the \HSC.
\end{itemize}

%%%%%%%%%
\section{Preliminaries}\label{sec:review}
%%%%%%%%%
 
 %%------------------
\subsection{Universality classes}\label{sec:class}
%%------------------

%%------------------
\subsubsection{Static universality class}%\label{sec:london}
%%------------------
As is well-known, a critical phenomenon is classified by a universality class. In a critical phenomenon, various quantities associated with the order parameter diverge at the critical point. The divergences are parametrized by critical exponents. These exponents define a universality class.

For example, the minimal \HSFHSC\ belongs to the $|\psi|^4$ mean-field theory universality class. The very first paper of \HSC\ pointed out that the condensate takes the value of the mean-field exponent, and this is further confirmed by computing all critical exponents \cite{Maeda:2009wv}.  In the holographic duality, one takes the large-$N_c$ limit so that fluctuations are suppressed, so one obtains mean-field theories.

Incidentally, when one says a mean-field theory, it implicitly assumes the $|\psi|^4$ theory. 
%\margin{E OK}
But we prefer to say the $|\psi|^4$ mean-field theory. This is because one may obtain mean-field theories other than the $|\psi|^4$ theory. The $\mathcal{N}=4$ super-Yang-Mills (SYM) theory at a finite chemical potential is an example (\appen{N=4}).

%%------------------
\subsubsection{Dynamic universality class}%\label{sec:class}
%%------------------

In dynamic critical phenomena, the relaxation time $\tau$ of the order parameter also diverges as $\tau \propto \xi^z$, which is known as the critical slowing down. The exponent $z$ is the dynamic critical exponent. The effect of slow dynamics is not limited to the order parameter. The other transport coefficients also have singular behaviors in general due to couplings with the order parameter.

The details of the dynamic exponent depend on dynamic universality classes. The dynamic universality classes were classified by Hohenberg and Halperin \cite{hohenberg_halperin}: they are known as model A, B, C, (D, E), F, G, H, and J.%
\footnote{Model D and E are written in parentheses since model D reduces to model B and model E is a special case of model F with an additional discrete symmetry.}
The classification is based on 
\begin{enumerate}
\item
whether the order parameter is conserved or not,
\item
whether there exist the other conserved charges,
\item 
% v2
whether there exist couplings between these modes (mode-mode coupling).
\end{enumerate}

Let us start with the simplest systems where only the order parameter matters near the critical point. They are model A and B: 
\begin{enumerate}
\item model A: the order parameter is not a conserved quantity ($z=2$).
\item model B: the order parameter is a conserved quantity ($z=4$).
\end{enumerate}
In field theories, $z$ can differ from the above values, but holographic results are the large-$N_c$ limit so that one obtains mean-field results. In order to obtain a nontrivial exponent, one needs to take $1/N_c$ corrections that are quantum gravity effects.

In the holographic duality, the study of dynamic universality class was initiated in Refs.~\cite{Maeda:2008hn,Maeda:2009wv} 
%based on the above ``traditional" classes:
focusing on the high-temperature phase: 
\begin{itemize}
\item Model A: A \HSFHSC\ belongs to this class because the order parameter is the ``macroscopic wave function" $\psi$ that is not conserved. 
The dynamic exponent is $z=2$ that is consistent with model A \cite{Maeda:2009wv}.
\item Model B: The $\mathcal{N}=4$ SYM at a finite chemical potential belongs to this class because the order parameter is a charge density \cite{Maeda:2008hn}. The critical exponent $z=4$ was shown explicitly in Ref.~\cite{Buchel:2010gd}. The dynamic exponent is $z=4$ that is consistent with model B.
\end{itemize}

For the other dynamic universality classes, the order parameter is coupled to the other modes. As the result of the mode-mode coupling, a dynamic critical exponent is typically modified. For example,
\begin{enumerate}
\item
model F: a non-conserved order parameter is coupled to the other conserved modes.
\item
model H: a conserved order parameter is coupled to the other conserved modes.
\end{enumerate}
A superfluid belongs to model F. For a \HSFHSC, $\psi$ couples to the conserved charge density $\rho$ in the low-temperature phase. Thus, one expects that a \HSFHSC\ belongs to model F even in the probe limit. 
 
In the context of gauge theories, the study of dynamic universality class was initiated in Ref.~\cite{Rajagopal:1992qz}. It became particularly popular after the work of Son and Stephanov \cite{Son:2004iv}: they identified that QCD belongs to model H universality class. %In model H, the conserved order parameter is coupled to the other conserved quantities.

Similarly, the $\mathcal{N}=4$ SYM at a finite chemical potential should belong to model H universality class \cite{Natsuume:2010bs}. %The $\mathcal{N}=4$ SYM undergoes a second-order phase transition at a finite chemical potential, and a charge shows a singular behavior at the critical point, so the charge corresponds to the order parameter. 
The charge is coupled to the other conserved charge, energy-momentum tensor $T^{\mu\nu}$, which implies that the theory actually belongs to model H. However, the model H effect has $1/N_c$ suppressions so that the model H effect is not visible in the holographic duality. %Namely, the dynamic exponent does not change from model B.

%The study of the other universality classes was initiated in Ref.~\cite{Natsuume:2010bs} which argued that the $\mathcal{N}=4$ SYM must belong the model H like QCD. 

%%------------------
\subsection{The holographic superfluid/superconductor}%\label{sec:background}
%%------------------

We consider the bulk 5-dimensional ``minimal" \HSFHSC:%
\footnote{We use upper-case Latin indices $M, N, \ldots$ for the 5-dimensional bulk spacetime coordinates and use Greek indices $\mu, \nu, \ldots$ for the 4-dimensional boundary coordinates. The boundary coordinates are written as  $x^\mu = (t, x^i) =(t, \vecx)=(t,x,y,z)$. $\omega$ and $q$ are collectively written as $k_\mu=(\omega,q,0,0)$.
}
\begin{subequations}
%\label{eq:}
\begin{align}
S_\text{bulk} &= \frac{1}{16\pi G_5}\int d^5x \sqrt{-g}(R-2\Lambda)+S_\text{m}~, \\
S_\text{m} &= -\frac{1}{g^2L} \int d^5x \sqrt{-g} \biggl\{ \frac{1}{4}F_{MN}^2 + L^2(|D_M\Psi|^2+m^2 |\Psi|^2) \biggr\}~, \\
F_{MN} &=\del_M A_N -\del_N A_M~, \quad
D_M =\nabla_M-iA_M~,\quad
\Lambda =-\frac{6}{L^2}~.
%\quad
%\Lambda =-\frac{6}{L^2}~.
%\label{eq:}
%
\end{align}
\end{subequations}
Here, we choose the mass dimensions as $[A_M]=\mass$, $[\Psi]=\mass^2$, and $[g]=\mass^0$. 
%We set $g=1$ below.
%The coupling $g$ has the dimension $[g]=\mass$, but we set $g=1$ below. 
We take the probe limit where the backreaction of the matter fields onto the geometry is ignored:
\begin{align}
\frac{1}{g^2N_c^2} \ll 1~,
%\label{eq:}
%
\end{align}
where $N_c$ is the number of ``colors" $N_c^2=8\pi^2L^3/(16\pi G_5)$, namely the probe limit corresponds to  the large-$N_c$ limit.
We set the bulk scalar charge $g=1$ below for simplicity.
Then, the background metric is given by the Schwarzschild-AdS$_5$ (SAdS$_5$) black hole:
\begin{subequations}
\label{eq:sads5}
\begin{align}
ds_5^2 &= \left( \frac{r}{L} \right)^2(-fdt^2+dx^2+dy^2+dz^2)+L^2\frac{dr^2}{r^2f} \\
&= \left( \frac{r_0}{L} \right)^2\frac{1}{u} (-fdt^2+dx^2+dy^2+dz^2)+L^2\frac{du^2}{4u^2f}~, \\
f &= 1-\left(\frac{r_0}{r}\right)^4 = 1-u^2~, 
%\label{eq:}
%
\end{align}
\end{subequations}
where $u:=r_0^2/r^2$.
The Hawking temperature is given by $\pi T =r_0/L^2$. For simplicity, we set the AdS radius $L=1$ and the horizon radius $r_0=1$, so we work in the unit $\pi T=1$.
The bulk matter equations are given by
\begin{subequations}
%\label{eq:}
\begin{align}
0 &= D^2\Psi-m^2\Psi~, \\
0 &= \nabla_NF^{MN} - J^M~,\\
J_M &= -i\{ \Psi^* D_M\Psi -\Psi(D_M\Psi)^*\} 
= 2\Im(\Psi^* D_M\Psi)~.
%\label{eq:}
%
\end{align}
\end{subequations}
In this paper, we consider the bulk scalar mass with $m^2=-4$. 
In the $A_u=0$ gauge, the $u\to 0$ asymptotic behaviors of matter fields are given by
%\margin{?}
\begin{subequations}
\label{eq:dictionary1}
\begin{align}
A_\mu &\sim \calA_\mu + A_\mu^{(+)} u~, \\
\Psi &\sim \frac{J}{2} u\ln u-\psi u~.
%\label{eq:dictionary1}
%
\end{align}
\end{subequations}
$\psi$ is the order parameter and $J$ is the source of the order parameter. 
% v2
Similarly, $\calA_t=\mu$ is the chemical potential, and $\calA_i$ is the vector potential. 
%$\calJ^t$ is the charge density, and $\calJ^i$ is the current density.  
According to the standard AdS/CFT dictionary, the currents are given by
\begin{subequations}
%\label{eq:}
\begin{align}
\bra\calJ^t \ket &= -\sqrt{-g}F^{tu}|_{u=0}~,\\
\bra\calJ^x \ket &= -\sqrt{-g}F^{xu}|_{u=0}~.
%\label{eq:}
%
\end{align}
\end{subequations}
Below we often omit brakets for simplicity.
In the bulk 5-dimensions, one needs the counterterm action for the Maxwell field to cancel UV divergences.
%, but the counterterm makes no contribution at the leading order. 

%According to the standard AdS/CFT dictionary,
%\begin{subequations}
%\label{eq:dict}
%\begin{align}
%
%\Psi &\sim \frac{J}{2} u\ln u-\psi u~,\\
%A_t &\sim \calA_t - \frac{\calJ^t}{2} u~,\\
%A_i &\sim \calA_i + \frac{\calJ^i}{2} u~,
%\label{eq:}
%
%\end{align}
%\end{subequations}

We consider perturbations of the form $e^{-i\omega t+iqx}$. 
We consider perturbations around the background:
\begin{subequations}
%\label{eq:}
\begin{align}
\Psi &= \bmPsi+\delta\Psi~, \\
A_t &= \bmA_t+a_t~, \\
A_i &= 0 +a_i~,
%\label{eq:}
%
\end{align}
\end{subequations}
where boldface letters indicate the background solution.

At high temperature, the background solution is given by
\begin{align}
\bmA_t = \mu(1-u)~, \quad
\bmA_i = 0~, \quad
\bmPsi =0~. 
\label{eq:sol_high}
\end{align}
Here, $\mu$ is the chemical potential. A \HSFHSC\ has 2 dimensionful control parameters $T$ and $\mu$, so the system is parameterized by a dimensionless parameter $\mu/T$. 
One can fix $T$ and vary $\mu$, or one can fix $\mu$ and vary $T$. We work in the unit $\pi T=1$, so we fix $T$.

The $\Psi=0$ solution becomes unstable at the critical point and is replaced by a $\Psi\neq0$ solution. When the background is the SAdS$_5$ black hole,  there exists a simple analytic solution at the critical point for $m^2=-4$ \cite{Herzog:2010vz}:
\begin{align}
\Psi \propto -\frac{u}{1+u}~,  \quad\text{at}\quad \left(\frac{\mu}{\pi T} \right)_c =2~.%\Delta=2~.
%\label{eq:herzog}
%
\end{align}
Below we utilize this solution to explore the system.

\paragraph{Counterterms:}
In the bulk 5-dimensions, one needs the counterterm action for the Maxwell field to cancel UV divergences:
\begin{align}
S_\text{CT}= -\int d^4x\, 
\frac{1}{4}
%\frac{1}{4g^2}
\sqrt{-\gamma}\gamma^{\mu\nu}\gamma^{\rho\sigma}F_{\mu\rho}F_{\nu\sigma} \times \ln(u^{1/2}/r_0)~,
%\label{eq:CT}
%
\end{align}
where $\gamma_{\mu\nu}$ is the 4-dimensional boundary metric (the 4-dimensional part of the bulk metric). The log term takes the form $\ln\tilde{u}$ if one uses $\tilde{u}:=L/r$. We use $u=(r_0/r)^2=r_0^2\tilde{u}$, so $\ln\tilde{u}=\ln(u^{1/2}/r_0)$. 
%Then, one obtains 
%\begin{align}
%
%\bra \calJ^\mu \ket = \left. 2\del_u A_\mu
%- \del_\nu(\sqrt{-\gamma}F^{\mu\nu}) \times \NGB^2 \ln(u^{1/2}/r_0) \right|_{u=0}~.
%\label{eq:}
%
%\end{align}
Then,
\begin{align}
\bra \calJ^\mu \ket = \left. 2\del_u A_\mu
- \del_\nu(\sqrt{-\gamma}F^{\mu\nu}) \times\ln(u^{1/2}/r_0) \right|_{u=0}~.
%\label{eq:}
%
\end{align}
At the leading order, the counterterm is necessary only for the Maxwell vector mode $\bra\calJ^y\ket$. 
For $A_y\propto e^{-i\omega t+iqx}$,
\begin{align}
\bra \calJ^y \ket = \left. 
%\frac{2}{g^2}
2\del_u A_y
+ 
%\frac{1}{g^2}
(\omega^2-q^2)\calA_y\left(\half\ln u-\ln r_0 \right) \right|_{u=0}~.
%\label{eq:current_dict}
%
\end{align}
We assume $O(\omega)\sim O(q^2)$, so the $\omega^2$-term does not contribute at the leading order.

% v2
\paragraph{Holographic \SC:}
So far, the boundary Maxwell field $\calA_\mu$ is added as an external source. To make the boundary Maxwell field dynamical, impose the ``holographic semiclassical equation"  as the boundary condition \cite{Natsuume:2022kic}:
\begin{align}
\del_\nu \calF^{\mu\nu} =e^2 \bra \calJ^\mu \ket~,
%\label{eq:semiclassical}
%
\end{align}
where $F_{\mu\nu} =\del_\mu \calA_\nu -\del_\nu \calA_\mu$. Here, all quantities including the $U(1)$ coupling $e$ are the boundary ones.

For example, consider the high-temperature phase, where the Maxwell perturbation $A_M$ decouples from the scalar perturbation $\delta\Psi$. The resulting theory can be regarded as the ${\mathcal N}=4$ SYM with dynamical photon. 

However, $e$ is the bare coupling constant, and it runs with scale. 
From the dual theory point of view, the running coupling constant $e_R$ comes from integrating out matter fields in the ${\mathcal N}=4$ SYM (see, \eg, Sec.~2 of Ref.~\cite{Fuini:2015hba}). From the holographic point of view, the running coupling constant comes from the holographic renormalization. 

However, this is the $T=0$ case. Our background is the SAdS$_5$ black hole, namely $T\neq0$, and the system lives in a medium, the ${\mathcal N}=4$ plasma. This breaks Lorentz invariance, and unlike the $T=0$ case, a single renormalized coupling constant $e_R$ is not enough. Instead, one needs to introduce the electric permittivity $\epse$ and the magnetic permeability $\mu_m$ (\sect{epse}).

%%%%%%%%%
\section{Bulk results: \HSF}\label{sec:bulk}
%%%%%%%%%

We consider linear perturbations of the form $e^{-i\omega t+iqx}$. We consider
\begin{itemize}
\item
The scalar perturbations $\delta\Psi, \delta\Psi^*$
\item
The Maxwell scalar perturbations $a_t,a_x,a_u$.
\item
The Maxwell vector perturbation $a_y$.
\end{itemize}
In order to obtain analytic solutions, 

\begin{itemize}
\item
Model F equations implicitly assume $O(|\psi|^2)\sim O(\del_i^2) \sim O(\del_t)$. To keep this counting in holographic computations, we set  
\begin{align}
\epsmu\to l^2 \epsmu, \eps \to l\eps, q \to l q, \omega \to l^2 \omega~,
%\label{eq:}
%
\end{align}
expand various fields in terms of $l$, and solve field equations at each order. As an example, see \sect{high} for the $\delta\Psi$-perturbation in high-temperature phase.

\item
We assign $a_x\sim O(l), a_t\sim O(l^2)$ so that the gauge-invariant combination is $O(l^2)$:
% old 
% We assign $a_x\sim O(l^0), a_t\sim O(l)$ so that the gauge-invariant combination is $O(l)$:
\begin{align}
\mfa_t:=a_t+\frac{\omega}{q}a_x~.
%\label{eq:}
%
\end{align}
We also assign $a_y\sim O(l)$.
% old
%We also assign $a_y\sim O(l)$.

\item
We impose 
%(1) 
the incoming wave boundary condition at the horizon $u=1$.
%, (2) $\mfa_t|_{u=0}=\mfat$, $a_y|_{u=0}=\calA_y$.

\end{itemize}

%%------------------
\subsection{High temperature phase}\label{sec:high}
%%------------------

%We consider linear perturbations of the form $e^{-i\omega t+iqx}$.
In the high-temperature phase, there is no condensate $\bmPsi=0$, and the perturbation $\delta\Psi$ decouples from Maxwell perturbations, and one can consider the $\delta\Psi$ perturbation and Maxwell perturbations separately.

%%------------------
%\subsubsection{Complex scalar}%\label{sec:}
%%------------------

%We consider the perturbation of the form $e^{-i\omega t+iqx}$. 
%When $\Psi=0$, $\delta A_t$ and $\delta A_i$ decouple from the $\delta\Psi$-equation. 
First, consider the $\delta\Psi$-perturbation.
This problem has been solved in Ref.~\cite{Natsuume:2018yrg}.
%\margin{OK}
Set $\epsmu\to l^2 \epsmu, q \to l q, \omega \to l^2 \omega$, and expand $\delta\Psi$ as a series in $l$:
% different ansatz
\begin{align}
\delta\Psi = (1-u^2)^{-i\omega/4} (l F_1 + l^3F_3+\cdots)~.
%\label{eq:}
%
\end{align}
We impose the boundary conditions (1) the incoming-wave boundary condition at the horizon (2) no fast falloff other than $F_1$. Namely, the order parameter $\delta\psi$ comes only from $F_1$.

The solution is given by
\begin{subequations}
%\label{eq:}
\begin{align}
F_1 &= -\Cone\frac{u}{1+u}~, \\
F_3 &= \Cone\frac{q^2-2\epsmu-(3+i)\omega}{8}\frac{u\ln u}{1+u} + \Cone\frac{(1-i)\omega+2\epsmu}{4} \frac{u\ln(1+u)}{1+u}~,
%\label{eq:}
%
\end{align}
\end{subequations}
so the asymptotic form with $l\to1$ is given by 
\begin{subequations}
\label{eq:QNM_high}
\begin{align}
\delta\Psi &\sim \frac{1}{8}\Cone\{q^2-2\epsmu-(1-3i)i\omega\} u\ln u- \Cone u+\cdots~, 
\quad(u\to0) \\
\to J & = \frac{1}{4}\{ q^2-2\epsmu-(1-3i)i\omega\}\Cone~. 
%\to \tau_0 &= \left. \frac{J}{-i\omega\delta\psi} \right|_{q=\epsmu=0} 
%\label{eq:}
%
\end{align}
\end{subequations}
%Setting $J=0$ gives the pole of the quasinormal mode (QNM). The pole has the following dispersion relation:
%\begin{align}
%
%(1-3i)i\omega = q^2-2\epsmu+\cdots~.
%\label{eq:}
%
%\end{align}

The detailed discussion of the dual GL theory is not the main purpose of this section, but if one focuses on the order parameter dynamics in the high-temperature phase, the critical dynamics is described by the TDGL equation or by the model A equation \cite{Natsuume:2018yrg}:
\begin{subequations}
%\label{eq:TDGL}
\begin{align}
\tau_0\del_t\psi &= - \frac{\delta F}{\delta \psi^*} \\
&= c_K D^2\psi+a\psi-b\psi|\psi|^2+J~,
%\label{eq:}
%
\end{align}
\end{subequations}
where $F=\int d^3x f$ and $f$ is given in \eq{GL4}. For the perturbation of the form $\psi\sim e^{-i\omega t+iqx}$, the TDGL equation is
\begin{subequations}
\label{eq:high_TDGL}
\begin{align}
J & =\{ c_Kq^2-a - i\omega \tau_0 \}\dpsi~. \\
%\tau_0\del_t\dpsi &=  \frac{1}{4} \del_i^2\dpsi+\half\epsmu\dpsi +J~,\\
\tau_0 &=\frac{1-3i}{4}~,\quad c_K=\frac{1}{4}~, \quad a=\half\epsmu~. 
%\to 0&=\frac{1}{4}\{q^2-(1-3i)i\omega-2\epsmu \} \dpsi~.
%\label{eq:}
%
\end{align}
\end{subequations}
The equation takes the form the diffusion equation $\omega=-iDq^2$ with dynamic exponent $z=2$, and $\tau_0$ plays the role of the relaxation time of the order parameter. But $\tau_0$ is complex, so $\dpsi$ shows the damped oscillation. The damped oscillation was first pointed out in Refs.~\cite{Amado:2009ts,Maeda:2009wv}.
%%------------------
%\subsubsection{Maxwell scalar mode}%\label{sec:}
%%------------------

For the bulk Maxwell field, it is the same as the pure Maxwell field problem in the SAdS$_5$ background. Namely, the problem corresponds to the $\mathcal{N}=4$ SYM with a boundary Maxwell field added as an external source. (See \sect{epse} for the $\mathcal{N}=4$ SYM with a dynamical boundary Maxwell field.)
This problem was solved long time ago \cite{Policastro:2002se}, and it has been solved many times, so we only quote the results. The results can be obtained by taking the condensate $\eps=0$ limit in the low-temperature phase (\sect{low}). 
%but we include it for completeness. 

%In order to discuss the time-dependence in the Maxwell sector, it is helpful first to consider the high-temperature phase. In this case, the system is the \Nfour\ SYM theory with dynamical photon. 

The Maxwell scalar mode consists of $a_t,a_x,a_u$. 
We impose (1) the incoming wave boundary condition at the horizon $u=1$,
% and $\mfa_t^{(i)}$ is regular at $u=1$, 
and (2) 
%the slow falloff of $\mfa_t$ comes only from $\mfa_t^{(1)}$and 
$\mfa_t|_{u=0}=\mfat$.
Here, $\mfat$ is the gauge-invariant combination of the boundary Maxwell field:
\begin{align}
\mfat:=\delta \calA_t+\frac{\omega}{q} \delta\calA_x = \frac{1}{iq}\calF_{xt} = \frac{1}{iq} E^x~.
%\label{eq:}
%
\end{align}

The charge density and the current density are given by
\begin{subequations}
%\label{eq:}
\begin{align}
 \bra \delta\calJ^t _n\ket
 &= \frac{2q^2}{q^2-2i\omega} \mfat~, 
 % old
% \quad O(l)~,
 \\
 \bra \delta\calJ^x_n \ket
&= \frac{2q\omega}{q^2-2i\omega} \mfat~.
% old
%\quad O(l^2)~.
%
\end{align}
\end{subequations}
The subscript ``$n$" denotes the ``normal current" that survives even when the condensate $\eps=0$.
The current is conserved:
\begin{align}
\del_\mu   \bra \delta\calJ^\mu_n \ket = 0~.
% old
%\quad O(l^3)~.
%\label{eq:}
%
\end{align}
Similarly, consider the Maxwell vector perturbation of the form $a_y e^{-i\omega t+iqx}$. $a_y$ is gauge-invariant by itself.
We impose (1) the incoming wave boundary condition at the horizon $u=1$, and (2) $a_y|_{u=0}=\calA_y$.
The current is given by
\begin{align}
 \bra \delta\calJ^y_n \ket = (i\omega+q^2 \ln r_0) \calA_y~.
 % old
%\quad O(l^2)~.
%\label{eq:}
%
\end{align}
The $(\ln r_0)$ term appears as the result of the holographic renormalization.% (\appen{vector}).

When $q=0$ in the $\calA_t=0$ gauge, the solutions reduce to
\begin{subequations}
%\label{eq:}
\begin{align}
\bra \delta \calJ^x_n \ket &=i\omega \calA_x~, \\
\bra \delta \calJ^y_n \ket &=i\omega \calA_y~.
%\label{eq:}
%
\end{align}
\end{subequations}
This is Ohm's law $J^i=i\omega\sigma_n(\omega)\calA_i$ with the conductivity $\sigma_n(\omega)=1+\cdots$.
%\margin{E OK}

%%------------------
\subsection{Low temperature phase}\label{sec:low}
%%------------------

%%------------------
\subsubsection{Background}%\label{sec:london}
%%------------------

The background solution $\bmPsi, \bmA_t$ has been obtained many times \cite{Herzog:2010vz,Natsuume:2018yrg,Natsuume:2022kic,first}, so we only quote the results.
Near the critical point, the scalar field remains small, and one can expand matter fields:
\begin{subequations}
%\label{eq:}
\begin{align}
\bmA_t(u) &= A_t^{(0)}+\epsilon^2 A_t^{(2)} + \cdots~,\\
\bmPsi(u) &= \epsilon\Psi^{(1)} +\cdots~. 
%\label{eq:}
%
\end{align}
\end{subequations}
The solutions are given by
\begin{subequations}
%\label{eq:}
\begin{align}
\bmPsi &= -\eps\frac{u}{1+u} +O(\eps^3)~, \\
\bmA_t &=(2+\eps^2\mu_2)(1-u) - \eps^2 \frac{u(1-u)}{4(1+u)} \\
& \sim (2+\eps^2\mu_2)- \left(2+\eps^2\mu_2+\frac{1}{4}\eps^2 \right)u~, 
\quad (u\to0)~.
%\label{eq:}
%
\end{align}
\end{subequations}
The critical point is $\mu_c=2$. For the spontaneous condensate, $\mu_2=1/24$. Then, the chemical potential and the charge density of the background are given by
\begin{subequations}
%\label{eq:}
\begin{align}
\mu &= \bmA_t |_{u=0}= 2+\eps^2 \mu_2~,\\
\calJ^t  &= \sqrt{-g} F^{ut}|_{u=0} =-2\bmA_t'|_{u=0} = 2\mu+\half \eps^2~. 
%\label{eq:}
%
\end{align}
\end{subequations}
Then, the condensate $\eps$ and $\epsmu:=\mu-\mu_c$  are related by
\begin{align}
\epsilon^2=24\epsmu+O(\epsmu^2)~.
%\label{eq:}
%
\end{align}

%%------------------
\subsubsection{Solution}%\label{sec:london}
%%------------------

In the low-temperature phase, the scalar modes $a_t,a_x,\delta\Psi,\delta\Psi^*$ are all coupled. 
Set $q \to l q,\omega\to l^2\omega,\epsilon\to l\epsilon,$ and expand fields as a series in $l$. 
%\margin{TBC}
We impose (1) the incoming wave boundary condition at the horizon $u=1$, and (2) $\mfa_t|_{u=0}=\mfat$,
%We impose the following boundary conditions:
%\begin{enumerate}
%\item The incoming-wave boundary conditions at the horizon.
%\item 
%$a_t|_{u=0}=\delta\calA_t$, $a_x|_{u=0}=\delta\calA_x$. 
%\end{enumerate}
%Here, $\mfa_t:=a_t+\omega a_x/q$ is the gauge-invariant combination and is $O(l)$. We take $a_x\sim O(l^0)$. Then, $a_t\sim O(l)$ and  the gauge-invariant combination is $O(l)$. 

At the leading order, we obtain%
\footnote{
The $J=\Jc=0$ result was obtained by Herzog \cite{Herzog:2010vz}. 
Note $\epsilon_\text{Herzog}^2 = 2\epsilon^2$. 
$\delta\calJ^t,\delta\calJ^x$ are written in terms of the gauge-invariant variable $\mfA_t$, but $\calA_x$ appears in $\delta\psi,\delta\psi^*$ because they are not gauge-invariant variables.}
\begin{subequations}
\label{eq:solution_mfAt}
\begin{align}
\delta\calJ^t &= 2q^2\frac{\calQ}{\calP} \mfA_t + 4\eps \frac{\calQ_{t}}{\calP} J + 4\eps \frac{\calQ_{t}^*}{\calP} \Jc~, 
% old
%\quad O(l)
\\
\delta\calJ^x &= 2q\omega \frac{\calQ}{\calP} \mfA_t+ q\eps \frac{\calQ_{x}}{\calP} J - q \eps \frac{\calQ_{x}^*}{\calP} \Jc~,
% old
%\quad O(l^2)
\\
\delta\psi &= \frac{\eps}{q}\delta\calA_x + 4\eps \frac{\calQ_{t}}{\calP} \mfA_t + 2\frac{\calQ_s}{\calP} J
-2\eps^2\frac{Q_c}{\calP} \Jc~,
% old
%\quad O(l^0)
\\
%-\frac{2\eps^2(2q^2-16i\omega+\eps^2)}{\calP} \Jc~,\\
\delta\psi^* &= -\frac{\eps}{q}\delta\calA_x + 4\eps \frac{\calQ_{t}^*}{\calP} \mfA_t 
-2\eps^2\frac{Q_c}{\calP} J
%-\frac{2\eps^2(2q^2-16i\omega+\eps^2)}{\calP} J  
+2\frac{\calQ_s^*}{\calP} \Jc~.
% old
%\quad O(l^0)
%\label{eq:}
%
\end{align}
\end{subequations}
%or
%\begin{subequations}
%\label{eq:bulk_Green}
%\begin{align}
%
%\delta\calJ^t &= 2q^2\frac{\calQ}{\calP} \calA_t + 2q\omega\frac{\calQ}{\calP} \calA_x+ 4\eps \frac{\calQ_{t}}{\calP} J + 4\eps \frac{\calQ_{t}^*}{\calP} \Jc~, 
% old
%\quad O(l)
%\\
%\delta\calJ^x &= 2q\omega \frac{\calQ}{\calP} \calA_t + 2\omega^2\frac{\calQ}{\calP} \calA_x+ q\eps \frac{\calQ_{x}}{\calP} J - q \eps \frac{\calQ_{x}^*}{\calP} \Jc~,
% old
%\quad O(l^2)
%\\
%\delta\psi &= 4\eps \frac{\calQ_{t}}{\calP} \calA_t + q\eps\frac{\calQ_x}{\calP} \calA_x + 2\frac{\calQ_s}{\calP} J
%-2\eps^2\frac{Q_c}{\calP} \Jc~,
% old
%\quad O(l^0)
%\\
%\delta\psi^* &= 4\eps \frac{\calQ_{t}^*}{\calP} \calA_t - q\omega\frac{\calQ_x^*}{\calP} \calA_x
%-2\eps^2\frac{Q_c}{\calP} J
%+2\frac{\calQ_s^*}{\calP} \Jc~.
% old
%\quad O(l^0)
%\label{eq:}
%
%\end{align}
%\end{subequations}
Here, the solution has a common denominator
\begin{align}
\calP &= 12q^6-48iq^4\omega-168q^2\omega^2+240i\omega^3+8q^4\epsilon^2-18iq^2\omega\epsilon^2-28\omega^2\epsilon^2+q^2\epsilon^4~,
%\label{eq:}
%
\end{align}
and there are 5 types of numerators:
\begin{subequations}
%\label{eq:}
\begin{align}
\calQ &= 12q^4 - 24iq^2\omega -120\omega^2 + 20q^2\eps^2 - 74i\omega \eps^2 + 7\eps^4~,
\\
\calQ_t &=6q^4+24(1-i)q^2\omega-60i\omega^2+3q^2\eps^2+7\omega\eps^2~,
\\
\calQ_x &=12q^4 + 24(1-2i)q^2\omega - (72+96i)\omega^2 + 8q^2\eps^2 + (12-18i)\omega\eps^2 + \eps^4~,
\\
\calQ_s &=24 q^4 + 72(1-i)q^2\omega- (48+144i)\omega^2 + 14q^2\eps^2 + (24-16i)\omega\eps^2 + \eps^4~,
\\
\calQ_c &=2q^2-16i\omega+\eps^2~.
%\label{eq:}
%
\end{align}
\end{subequations}
We work in the momentum space (Fourier space). So, a complex conjugate solution is obtained by $\omega\to-\omega, q\to-q$ and taking the complex conjugate. $\omega$ appear in the form of $w := i\omega$, so take the complex conjugate as if $w$ is real.%
\footnote{We work in the momentum space, so $\psi$ here is $\psi_k$, where $\psi(t,\bmx,u)=\int (dk) e^{-i\omega t+i \bmq\cdot\bmx}\psi_k(u)$ with $(dk)=d^4k/(2\pi)^4$. Then, the complex conjugate field is given by $\psi_{-k}^*$.
}

The current is conserved when there are no scalar sources $J,\Jc$:
\begin{align}
\del_\mu\delta\calJ^\mu=i\epsilon (J-\Jc)~.
% old
%\quad O(l^3)~.
%\label{eq:}
%
\end{align}

\paragraph{Limits:}
\begin{itemize}
\item
The $\epsilon=0$ limit:
\begin{subequations}
%\label{eq:}
\begin{align}
\delta\calJ^t &= \frac{2q^2}{q^2-2i\omega} \mfA_t~, \\
\delta\calJ^x &= \frac{2q\omega}{q^2-2i\omega} \mfA_t~, \\
\delta\psi &= \frac{4}{q^2-(1-3i)i\omega}J~. 
%\label{eq:}
%
\end{align}
\end{subequations}
$\delta\calJ^t,\delta\calJ^x$ agree with the high-temperature result. $\dpsi$ agrees with the high-temperature result at the critical point. %Thus, the high-temperature results agree with the low-temperature results at the critical point. 
\item
The $\omega=0$ limit: 
\begin{subequations}
%\label{eq:}
\begin{align}
\delta\calJ^t &= \frac{1}{6q^2+\eps^2} \{2 (6q^2+7\eps^2) \delta\calA_t +12 \eps (J+\Jc) \} \\
&= 2 \delta\calA_t +\half\eps(\dpsi+\dpsi^*)~, \\
\delta\calJ^x &= \frac{\eps}{q} (J-\Jc), \\
\delta\psi &= \frac{12\eps}{6q^2+\eps^2} \delta\calA_t - \frac{\eps}{q}\delta\calA_x +\frac{24}{6q^2+\eps^2} \Jc -\frac{2\eps^2}{q^2(6q^2+\eps^2)} (J-\Jc)~.
%\delta\calJ^t &= \frac{2(6q^2+7\eps^2)}{6q^2+\eps^2}\calA_t +\frac{24\eps}{6q^2+\eps^2} J
%= 2\calA_t +\eps\dpsi~, \\
%\delta\calJ^x &=0~, \\
%\delta\psi &= \frac{12\eps}{6q^2+\eps^2}\calA_t +\frac{24}{6q^2+\eps^2} J~.
%\label{eq:}
%
\end{align}
\end{subequations}
In this case, one can set $\Cone^*=\Cone, J=\Jc, \delta\calA_x=0$: 
\begin{subequations}
%\label{eq:}
\begin{align}
\delta\calJ^t &= \frac{2(6q^2+7\eps^2)}{6q^2+\eps^2} \delta\calA_t +\frac{24\eps}{6q^2+\eps^2} J
= 2\delta\calA_t +\eps\dpsi~, \\
\delta\calJ^x &=0~, \\
\delta\psi &= \frac{12\eps}{6q^2+\eps^2}\delta\calA_t +\frac{24}{6q^2+\eps^2} J~.
%\label{eq:}
%
\end{align}
\end{subequations}
The case $\delta\calA_t=0$ has been obtained in Ref.~\cite{first}.
\end{itemize}

\paragraph{Pole structure:}
In the high-temperature phase, there are 3 independent scalar modes: $\delta\calJ^t,\Cone, \Cone^*$. In the low-temperature phase, they are no longer independent, but one expects 3 modes characterized by $\delta\calJ^t$ (mode M), $\Cone$ (mode S), and 
% v12
$\Cone^*$ (mode C). 

In fact, the solution has common poles at
\begin{align}
\calP &= 12q^6-48iq^4\omega-168q^2\omega^2+240i\omega^3+8q^4\epsilon^2-18iq^2\omega\epsilon^2-28\omega^2\epsilon^2+q^2\epsilon^4 =0~.
%\omega= -\half i q^2-\frac{1}{4}i \epsilon^2~.
\label{eq:pole}
\end{align}
The pole $\calP$ is an $O(\omega^3)$ expression, so there are three  $\omega\neq0$ solutions in general. They correspond to mode M, mode S, and mode C.
However, the exact expressions are rather involved. Schematically, we write
\begin{subequations}
%\label{eq:eps_small}
\begin{align}
\text{Mode M: }  \omega&=\omega_M(q,\eps) \to  \delta\calJ^t~. \\
\text{Mode S: } \omega&=\omega_S(q,\eps)  \to \delta\psi~. \\
\text{Mode C: } \omega&=\omega_C(q,\eps) \to \delta\psi^*~. 
%\label{eq:}
%
\end{align}
\end{subequations}
If one uses $w:=i\omega$, $\calP$ is a cubic equation in $w$ with real coefficients, so there is at least one real solution $w$. This corresponds to the Maxwell mode M. Namely, the pole of the Maxwell mode lies on the pure imaginary axis in the $\omega$-plane.

One can obtain these poles approximately. This was first analyzed in Ref.~\cite{Herzog:2010vz}.
%\footnote{Note $\epsilon_\text{Herzog}^2 = 2\epsilon^2$. }
When $\eps \ll q$, 
\begin{subequations}
\label{eq:eps_small}
\begin{align}
\omega &= -\frac{i}{2} q^2-\frac{i}{10} \epsilon^2 +\cdots~,\quad(\text{M})~. \\
\omega &= \frac{3-i}{10}q^2+\frac{9-i}{120}\epsilon^2 +\cdots~, \quad(\text{S})~. \\
\omega &= \frac{-3-i}{10}q^2+\frac{-9-i}{120}\epsilon^2 +\cdots~, \quad(\text{C})~. 
%\label{eq:}
%
\end{align}
\end{subequations}
The first pole corresponds to the diffusive pole with an $\epsilon$-correction.
The last two poles correspond to the ones for order parameters.

Similarly, when $\eps \gg q $, 
\begin{subequations}
\label{eq:q_small}
\begin{align}
\omega &= -\frac{7i}{60}\epsilon^2-\frac{89}{245}iq^2+\cdots~,
 \quad(\text{M})~.\\
 \omega &=+\frac{q\epsilon}{2\sqrt{7}}-\frac{33i}{196}q^2+\cdots~
 \quad(\text{S})~. \\
\omega &=-\frac{q\epsilon}{2\sqrt{7}}-\frac{33i}{196}q^2+\cdots~,
 \quad(\text{C})~.
%\label{eq:}
%
\end{align}
\end{subequations}
The last two poles have the linear dispersion and are proportional to $\epsilon$, so they are interpreted as the second sounds. The second terms of these poles represent the sound attenuation.

As one increases $\eps$, poles in Mode M,S, and C \eqref{eq:eps_small} become \eq{q_small}.

Finally, for the vector mode $a_y$, one obtains
\begin{subequations}
%\label{eq:}
\begin{align}
\delta \calJ^y 
&=\left(i\omega+ q^2\ln r_0 - \half \epsilon^2+\cdots \right)\calA_y~, 
% old
%\quad O(l^2)
\\
\sigma(\omega) &= \left. \frac{\delta \calJ^y }{i\omega\calA_y} \right|_{q=0} 
= \frac{i\epsilon^2}{2\omega}+1+\cdots~.
%\label{eq:}
%
\end{align}
\end{subequations}
$\text{Im}(\sigma)$ has the $1/\omega$-pole which implies the diverging DC conductivity.

%%%%%%%%%
\section{The dual model F}\label{sec:F}
%%%%%%%%%

%%------------------
%\subsection{The TDGL equation}\label{sec:TDGL}
%%------------------

%Again, this TDGL equation should be regarded as leading terms in the effective theory expansion. The purpose of this paper is to identify the dual TDGL equation at the next order in the effective theory expansion.

%Note that the TDGL equation implicitly assumes $O(|\psi|^2)\sim O(\del_i^2) \sim O(\del_t)$. This is appropriate from holographic results as we see below. %We keep this counting when we extend the TDGL equation at next order.

%%------------------
%\subsection{The model F equations}%\label{sec:}
%%------------------

We would like to identify the dual GL theory from bulk results in \sect{bulk}.
The scalar mode can be explained by model F equations.
Our model F equations are given by 
\begin{subequations}
\label{eq:F}
\begin{align}
D_t\psi &= - \tau_0^{-1}\frac{\delta F}{\delta \psi^*}  -ig_0 \psi\frac{\delta F}{\delta \rho}~, \\
\del_t \rho &= \lambda \del_i^2 \left( \frac{\delta F}{\delta \rho} \right) - \lambda \del_iE^i  +2g_0 \Im \left( \psi^*\frac{\delta F}{\delta \psi^*} \right)~,
%\label{eq:}
%
\end{align}
\end{subequations}
as well as the complex conjugate equation for $\psi^*$. 
Here, $D_\mu=\del_\mu-ig_0 \calA_\mu$ and $E^i$ is the electric field $E^i=\calF_{it} =\del_i \calA_t- \del_t \calA_i$. The GL free energy $F=\int d^3x f$ is given by
\begin{align}
f &= c_K|D_i\psi|^2 - a |\psi|^2 + \frac{b'}{2}|\psi|^4+
\frac{1}{2C}\rho^2 +\gamma \rho |\psi|^2
%-\rho \At 
-(\psi J^*+\psi^* J)~. 
\end{align}
We identify $\rho$ as the charge density and $\calA_t$ as the gauge potential. The $\rho$-equation represents the current conservation law.
% ($\rho$ in this section is not gauge-invariant variable).

Because we identify model F as the one with Maxwell field, we slightly modify model F equations so that the system includes both $\calA_t$ and $\calA_i$. We also write down the equations so that the $U(1)$ gauge symmetry is manifest:
\begin{enumerate}
\item
For the standard model F, the free energy has the $\rho \calA_t$ coupling and becomes the covariant derivative $D_t$. We omit this term from $f$ and include it in the equation of motion. 
\item
More importantly, the standard $\rho$-equation has only $\calA_t$ and does not couple to $\calA_i$. (For a \HSC, the vector potential should exist. For a holographic superfluid, it corresponds to a superfluid flow.) As a result, the conventional model F equation is not gauge-invariant and is valid only when $\calA_i=0$ (\appen{conventional}).
\end{enumerate}
%Namely, we prefer that the dissipative term, which breaks the time reversal symmetry, are added on the left-hand side of equations. 
%\margin{E}

% v2
Eqs.~\eqref{eq:F}  become
\begin{subequations}
%\label{eq:}
\begin{align}
D_t\psi &= \tau_0^{-1}[ c_K D_i^2\psi + a\psi - b'\psi|\psi|^2 - \gamma \rho\psi +J] -ig_0 \psi\left( \frac{1}{C} \rho + \gamma|\psi|^2\right)~, \\
%\tau_0 \del_t\psi &= c_K \del_i^2\psi+ \left(a-\gamma m \right)\psi-\frac{ig}{C}m\psi-(b+ig\gamma)\psi|\psi|^2 +J + ig\psi h \\
\del_t \rho &= \lambda \del_i^2 \left( \frac{1}{C} \rho + \gamma|\psi|^2 - \mfA_t \right) +2g_0 \Im \left( \psi^*\frac{\delta F}{\delta \psi^*} \right)~.
%\label{eq:}
%
\end{align}
\end{subequations}
% v12
Here,
\margin{TBC}
\begin{align}
2 \Im \left( \psi^*\frac{\delta F}{\delta \psi^*} \right)
= ic_K \{\psi^*D^2\psi-\psi (D^2\psi)^*\} - i(\psi \Jc-\psi^* J)~.
%\label{eq:}
%
\end{align}
The combination $D:=\lambda/C$ represents the diffusion constant for $\rho$. Similarly, $\tau_0$ represents the relaxation time for $\psi$, but $\tau_0 \in \mathbb{C}$. 

Comparing with our holographic results, we obtain
\begin{align}
\tau_0=\frac{1-3i}{4},
\quad
D=\half, 
\quad
C=2, 
\quad
g_0=1,
\quad
\gamma=-\frac{1}{4},
\quad
a=\half \epsmu,
\quad
b'
%=\frac{1}{8}+\frac{1}{48}
=\frac{7}{48}~.
\label{eq:F_parameters}
\end{align}

\begin{itemize}
\item
%The factor $ig\calA_t$ plays the role of covariant derivative. 
Because we use the bulk covariant derivative $D_M=\nabla_M-iA_M$, $g_0=1$.

\item
The mass dimensions are
\begin{subequations}
%\label{eq:}
\begin{align}
[\psi] &=\mass, \ [\calA_\mu] = \mass,\ [\rho]=[\calJ^i]=[J]=\mass^3, \\
[\tau_0]&=\mass,\ [D]=\mass^{-1}, \ [C]=\mass^2,  \ [\lambda]=\mass, \ [g_0]=\mass^0, \ [\gamma]=\mass^{-1}~,
%\label{eq:}
%
\end{align}
\end{subequations}
so that the free energy $f$ has $[f]=\mass^4$.
The parameter 
% v2
$C:=\del\rho/\del\mu$ is the charge susceptibility, and $\lambda$ is the conductivity.
Because we work in the unit $\pi T=1$, multiply appropriate powers of $(\pi T)$ to obtain the correct dimensions.

%\item
%The readers may notice that the parameter $b'$ are different from the one that is determined previously \cite{Natsuume:2018yrg,Natsuume:2022kic,first}:
%\begin{align}
%
%b=\frac{1}{48}~.
%\label{eq:}
%
%\end{align}
%\margin{E}
%But in the stationary case, model F equations reduce to the free energy (\sect{F_to_GL}). 
%and it is convenient to absorb the $\rho|\psi|^2$ term into $b$. 
%To avoid the confusion, we use the convention $b'$ here.
\end{itemize}

%%------------------
%\subsection{Determining model F parameters}%\label{sec:}
%%------------------

\paragraph{High-temperature phase:}
Consider the linear perturbations. In this case, $\eps=0$, and the $\delta\psi$-perturbation and $\delta\rho$-perturbation decouple. The $\delta \rho$-equation becomes the diffusion equation:
\begin{subequations}
%\label{eq:}
\begin{align}
0&=\del_t \delta\rho-D\del_i^2\delta(\rho -C \mfA_t) \\
&=( -i\omega +Dq^2)\delta\rho-C Dq^2\mfA_t~, \\
\to \delta\rho&= \frac{CD q^2}{Dq^2-i\omega}\mfA_t~.
%\label{eq:}
%
\end{align}
\end{subequations}
The holographic result is 
\begin{subequations}
%\label{eq:}
\begin{align}
\delta\calJ^t_n &= \frac{2q^2}{q^2-2i\omega} \mfA_t~, \\
\to D&=1/2, C=2~.
%\label{eq:}
%
\end{align}
\end{subequations}
\begin{enumerate}
\item
When $\omega=0$, $\delta\rho=C\delta\mu$. This is the constant chemical potential case. For \HSFHSCs, one always uses the solution $\delta\calA_t=\delta\mu(1-u)$, so one always has $C=2$ for any \HSFHSCs\ in the SAdS$_5$ background. Similarly, for the SAdS$_4$ background, $C=4\pi T/3$. 
% v2
The charge susceptibility is given by
\begin{align}
\chi_T = \frac{\del\rho}{\del \mu}=C~.
%\label{eq:}
%
\end{align}

\item
The $\delta\rho$-equation is the conservation law, so the current is given by
\begin{align}
\delta\calJ^x_n =CD \frac{q\omega}{Dq^2-i\omega}\mfA_t~.
%\label{eq:}
%
\end{align}
When $q=0$,
\begin{align}
\delta\calJ^x_n = \lambda E^x. %i\omega \lambda\delta\calA_x~.
%\label{eq:}
%
\end{align}
This is Ohm's law $\delta\calJ^x_n=\sigma_n(\omega)E^x$, so $\lambda$ is the conductivity. Then, the Einstein relation holds:
\begin{align}
\sigma =D\chi_T\to \lambda=CD~.
%\label{eq:}
%
\end{align}
\end{enumerate}

The $\dpsi$-equation is given by
\begin{subequations}
%\label{eq:}
\begin{align}
\del_t\dpsi &= \tau_0^{-1}[ c_K \del_i^2\dpsi+a\dpsi +J]~,\\
\to J&=\frac{1}{4}\{q^2-(1-3i)i\omega-2\epsmu \} \dpsi~.
%\label{eq:}
%
\end{align}
\end{subequations}
Then, the $\dpsi$-equation reduces to the TDGL equation \eqref{eq:high_TDGL}.
%Here, we use the result $\gamma=-1/4$ obtained below.

\paragraph{Low-temperature phase:}
From the $\psi$-equation, the background $\eps, \rho_0$ is given by ($J=0$)
\begin{align}
0= -ig\eps \left(\frac{\rho_0}{C}+\gamma\eps^2 \right)+\eps\frac{a-\rho_0\gamma-b'\eps^2}{\tau_0}
%\label{eq:}
%
\end{align}
The $\psi^*$ equation gives the complex conjugate equation, and $\rho_0$ cannot vanish. The real part and the imaginary part must vanish separately:
%\margin{TBC}
\begin{subequations}
%\label{eq:}
\begin{align}
0&=\frac{\rho_0}{C}+\gamma\eps^2 \to \rho_0 = - 2\gamma\eps^2 = \half\eps^2 \\
&\to \gamma=-\frac{1}{4} \\
0&=a - \rho_0\gamma - b'\eps^2 = \left(\frac{1}{48} + \frac{1}{8} -b' \right)\eps^2 \\
%a+(c\gamma^2-b)\eps^2 =a+ \left(\frac{1}{4}-b \right)\eps^2
&\to b' 
%=\frac{1}{48}+\frac{1}{8} 
= \frac{7}{48}
%\label{eq:}
%
\end{align}
\end{subequations}
Here, we use $\epsmu=\eps^2/24$. 
%Namely,
%\margin{TBC}
%\begin{subequations}
%\label{eq:}
%\begin{align}
%
%f &= \frac{1}{4}|\del_i\psi|^2 - \half \epsmu |\psi|^2 + \frac{1/48+1/4}{2}|\psi|^4+\cdots+
%\frac{1}{2}m^2 -\half m |\psi|^2-mh -(\psi J^*+\psi^* J)~,\\
%m_0&= 2\mu+\half \eps^2
%
%\end{align}
%\end{subequations}

%\paragraph{Low-temperature phase:}
Consider the linear perturbations $\dpsi,\dpsi^*,\delta \rho$ around the background. Solve $\dpsi,\dpsi^*,\delta \rho$ equations in terms of sources $\mfA_t, J, \Jc$. This reproduces bulk results \eqref{eq:solution_mfAt}.

\begin{subequations}
%\label{eq:}
The generic dispersion relations by model F parameters are rather complicated, but when $\eps\gg q$,
\begin{align}
\omega &=-i\frac{ (\tau_0+\tau_0^*)(a+\gamma^2\chi_T \eps^2)}{\tau_0\tau_0^*}+\cdots~,\\
\omega &= \pm\sqrt{ \frac{2ac_K}{\chi_T (a+\gamma^2\chi_T \eps^2)} }q\eps+\cdots~.
%\label{eq:}
%
\end{align}
\end{subequations}
% v12
Here, $\tau_0^*$ is the complex conjugate of $\tau_0$.
This of course reproduces the bulk result \eqref{eq:q_small}.

%%%%%%%%%
\section{Bulk results: \HSC}%\label{sec:}
%%%%%%%%%

%%------------------
\subsection{The electric permittivity and magnetic permeability}\label{sec:epse}
%%------------------

So far, the boundary Maxwell field is added as an external source. To make the boundary Maxwell field dynamical, impose the ``holographic semiclassical equation"  as the boundary condition \cite{Natsuume:2022kic}:
\begin{align}
\del_\nu \calF^{\mu\nu} =e^2 \bra \calJ^\mu \ket~.
\label{eq:semiclassical}
\end{align}
Here, all quantities including the $U(1)$ coupling constant $e$ are the boundary ones.
The vacuum magnetic permeability $\mu_0$ and the vacuum electric permittivity $\varepsilon_0$ are given by $\mu_0=1/\varepsilon_0=e^2$. 

However, \HSCs\ do not live in the vacuum: instead they live in a medium, a large-$N_c$ plasma. This breaks Lorentz invariance, and unlike the vacuum case, a single coupling constant $e^2$ is not enough. Just like elementary electrodynamics, one needs to introduce $\varepsilon_e$ and $\mu_m$, and they differ from the vacuum values. In other words, we should consider the macroscopic Maxwell equation. 

In this case, the medium effect comes from the normal current that survives even when the condensate $\eps=0$, so it is enough to consider the high-temperature phase. The dual theory can be regarded as the $\mathcal{N}=4$ SYM with a dynamical photon, and we compute $\varepsilon_e$ and $\mu_m$ for the photon.

We decompose the current as the normal current and the rest (such as the external current $\calJ_\text{ext}^\mu$ or the supercurrent) $\calJ^\mu=\calJ_n^\mu+\calJ_\text{ext}^\mu$. 
We make the tensor decomposition for the normal current:
\begin{subequations}
%\label{eq:}
\begin{align}
\bra \calJ_n^t \ket &= -\frac{q^2}{\omega^2} K_L \mfA_t~, \\
\bra \calJ_n^x  \ket &= -\frac{q}{\omega} K_L \mfA_t~, \\
\bra \calJ_n^y  \ket &= -K_T \calA_y~.
%\label{eq:}
%
\end{align}
\end{subequations}
Then, impose the holographic semiclassical equation: 
\begin{subequations}
%\label{eq:}
\begin{align}
\del_i\calF^{ti} &= e^2 \bra\calJ_n^t \ket +e^2\calJ^t_\text{ext}~, \\
\to -\frac{1}{e^2}q^2 \mfat &= -\frac{q^2}{\omega^2}K_L\mfat +\calJ^t_\text{ext}~, \\
\to-\epse(k) q^2  \mfat &= \calJ^t_\text{ext}~.
%\\
%\epse &:=\frac{1}{e^2} -\frac{K_L}{\omega^2}~.
%\label{eq:}
%
\end{align}
\end{subequations}
Here, $\calF_{tx}=-iq\mfat$, and $\epse(k)$ is the electric permittivity (dielectric constant):
%\margin{sign corrected}
\begin{subequations}
\label{eq:epse}
\begin{align}
\epse(k) &= \frac{1}{e^2} -\frac{K_L}{\omega^2} \\
& = \frac{1}{e^2}+\frac{2}{q^2-2i\omega} \\
& \sim\frac{2}{q^2-2i\omega}~, \quad O(l^{-2})
% \\
%& \stackrel{\omega\to0}{\longrightarrow} 
%\frac{2}{q^2}~.
%\label{eq:}
%
\end{align}
\end{subequations}
Thus, the net effect of the normal current is to change the electric permittivity from the vacuum value $\varepsilon=1/e^2$ to $\epse$. However, the bare electric permittivity $1/\varepsilon_0=e^2$ does not arise at the leading order in the $l$-expansion. 

Similarly, the $x$-component of the holographic semiclassical equation becomes
%\margin{sign?}
\begin{subequations}
%\label{eq:}
\begin{align}
\del_\nu\calF^{x\nu} &= e^2 \bra\calJ^x_n \ket +\calJ^x_\text{ext}~,\\
\to -\frac{1}{e^2}\omega q \mfat &=  -\frac{q}{\omega}K_L \mfat+\calJ^x_\text{ext}~. \\
\to -\omega q \epse(k) \mfat &= \calJ^x_\text{ext}~.
%\label{eq:}
%
\end{align}
\end{subequations}
In the limit $q\to0$, the equation reduces to Ohm's law:
%\margin{TBC}
\begin{align}
i\omega \epse(\omega)E^x = \calJ^x_\text{ext}
\to \epse(\omega) =\frac{\sigma_n(\omega)}{i\omega}~.
%\label{eq:}
%
\end{align}
In our case, $\sigma_n(\omega)=1+\cdots$.

For the vector mode $\calJ^y_n$, 
\begin{subequations}
%\label{eq:}
\begin{align}
%
%a_y & \sim \left\{ 1-\frac{1}{4}(q^2-2i\omega) u +\frac{1}{4} q^2 u\ln u+\cdots \right\} \calA_y~, \\
\bra \calJ^y_n \ket 
%&= \left. 2\del_u a_y
%+(\omega^2-q^2)\calA_y \left( \half\ln u-\ln r_0 \right) \right|_{u=0} 
%\\
&=( i\omega+q^2\ln r_0 ) \calA_y~, 
% old
%\quad O(l^2)
\\
&=-K_T \calA_y~, \\
K_T &=-(i\omega+q^2\ln r_0)~.
% old
%\quad O(l^2)~.
%&= -\frac{\epsilon^2}{2i\omega}+1+\cdots~.
%\label{eq:}
%
\end{align}
\end{subequations}
Then, the holographic semiclassical equation becomes
\begin{subequations}
%\label{eq:}
\begin{align}
\del_\nu \calF^{y\nu} &= e^2 \bra \calJ^y \ket \\
(-\omega^2+q^2) \calA_y &= -e^2K_T\calA_y+ e^2 \calJ^y_\text{ext}~, \\
 \calJ^y_\text{ext} &= \left( \frac{\cdots+q^2}{e^2} + K_T \right)\calA_y \\
 &=-\omega^2 \left(\cdots - \frac{K_L}{\omega^2} \right)\calA_y 
 + \frac{1}{e^2} \left(1+e^2 \frac{K_T-K_L}{q^2} \right)q^2 \calA_y \\
 &= \left(-\epse(k)\omega^2+\frac{1}{\mu_m(k)} q^2 \right) \calA_y~.
 % cancel
%\del_\nu \calF^{y\nu} &= e^2 \bra \calJ^y \ket \\
%(-\cancel{\omega^2}+q^2) \calA_y &= -e^2K_T\calA_y+ e^2 \calJ^y_\text{ext}~, \\
% \calJ^y_\text{ext} &= \left( \frac{\cancel{-\omega^2}+q^2}{e^2} + K_T \right)\calA_y \\
% &=-\omega^2 \left(\cancel{\frac{1}{e^2}} - \frac{K_L}{\omega^2} \right)\calA_y 
% + \frac{1}{e^2} \left(1+e^2 \frac{K_T-K_L}{q^2} \right)q^2 \calA_y \\
% &= \left(-\epse(k)\omega^2+\frac{1}{\mu_m(k)} q^2 \right) \calA_y~.
%\label{eq:}
%
\end{align}
\end{subequations}
Here, $\cdots$ represents $O(\omega^2)$ terms that do not contribute at the leading order because $\omega^2=O(l^4)$.
%We define $\varepsilon_e$ so that  $\calJ^x_\text{ext}, \calJ^y_\text{ext}$ take the same form.
At the leading order, $K_L, K_T=O(l^2)$.
%Note that the Maxwell equation contains the $\omega^2=O(l^4)$ term, but this is higher order in the $l$-expansion.
%However, at the leading order, $K_L, K_T=O(l^2)$ so that the above semiclassical equation is not valid. The LHS contains the $\omega^2=O(l^4)$ term. 
%We have to go to next order.
The magnetic permeability then becomes
%\margin{?corrected Jan.2025}
\begin{subequations}
%\label{eq:}
\begin{align}
\frac{1}{\mu_m(k)} &= \frac{1}{e^2} + \frac{K_T-K_L}{q^2} \\
&\sim \frac{1}{e^2} -\ln r_0 - \frac{i\omega}{q^2-2i\omega}
% old
%\quad O(l^0) 
\\
& \stackrel{\omega\to0}{\longrightarrow} 
\frac{1}{e^2} -\ln r_0~.
%\label{eq:}
%
\end{align}
\end{subequations}
Then, the net effect of the normal current is to change the magnetic permeability from the vacuum value $\mu_0=e^2$ to $\mu_m$. When $\omega=0$, $\mu_m$ was obtained in Ref.~\cite{Natsuume:2022kic}.
%For $\mu_m>0$, $e^2\ln r_0<1$.

Then, we obtain the following set of the dual Maxwell equation for \HSCs:
\begin{subequations}
%\label{eq:}
\begin{align}
-\epse(k) q^2 \mfA_t &= \bra \calJ^t_s \ket~,\\
-\epse(k) q\omega \mfA_t &= \bra \calJ^x_s \ket~,\\
\left(-\epse(k)\omega^2+\frac{1}{\mu_m(k)} q^2 \right) \calA_y &= \bra \calJ^y_s \ket~.
%\label{eq:}
%
\end{align}
\end{subequations}
Here, $\calJ^\mu_s=\calJ^\mu-\calJ^\mu_n$. In this way, all normal current contributions can be expressed by $\epse,\mu_m$.

%%------------------
\subsection{Scalar perturbations}%\label{sec:}
%%------------------

As was mentioned above, for the Maxwell scalar mode, the added kinetic term is not the leading order in the $l$-expansion. The semiclassical equation is
\begin{align}
-\frac{1}{e^2}q^2 \mfA_t = \bra\delta\calJ^t \ket+\calJ^t_\text{ext}~.
%\label{eq:}
%
\end{align}
The left-hand-side is $O(l^2) \mfA_t$ whereas the right-hand side is $O(l^0) \mfA_t$.  Then, the scalar mode problem reduces to the Neumann boundary condition problem at the leading order:
\begin{align}
0= \bra\delta\calJ^t \ket+\calJ^t_\text{ext}~.
%\label{eq:}
%
\end{align}
%For the scalar mode, the problem reduces to the Neumann boundary condition problem at the leading order.
For simplicity, we set $J=\Jc=0$. Using the solution \eqref{eq:solution_mfAt}, one obtains $\mfA_t$, and substituting it into $\dpsi$. In terms of gauge-invariant variables, one obtains
%\begin{align}
%
%\delta\calJ^t &=2q^2 \frac{\calQ}{\calP} \delta\mfA_t
%\to \delta\mfA_t = - \frac{1}{2q^2}\frac{\calP}{\calQ}\calJ^t_\text{ext}
%\label{eq:}
%
%\end{align}
%Substituting the result into $\dpsi$, one obtains
\begin{subequations}
%\label{eq:}
\begin{align}
\mfA_t &= - \frac{1}{2q^2}\frac{\calP}{\calQ}\calJ^t_\text{ext}~, \\
\mfA_x &= \frac{1}{q} \frac{\calQ_t - \calQ_t^*}{\calQ} \calJ^t_\text{ext}~, \\
\dpsi_+ &= -\frac{\eps}{q^2} \frac{\calQ_t + \calQ_t^*}{\calQ} \calJ^t_\text{ext}~, \\
\mfA_x &:= \delta \calA_x -\frac{iq}{\eps}\frac{\dpsi-\dpsi^*}{2i}~,\quad
\dpsi_+ := \frac{\dpsi+\dpsi^*}{2}~.
%\\
%\dpsi &= \frac{\eps}{q}\delta\calA_x -\frac{2\eps}{q^2} \frac{Q_t}{Q} \calJ^t_\text{ext}~, \\
%\dpsi^* &=-\frac{\eps}{q}\delta\calA_x - \frac{2\eps}{q^2} \frac{Q_t^*}{Q} \calJ^t_\text{ext}~.
 %\label{eq:}
%
\end{align}
\end{subequations}

The solution $\mfA_t, \mfA_x,\dpsi_+$ now has common poles at
 \begin{align}
\calQ &= 12q^4 - 24iq^2\omega -120\omega^2 + 20q^2\eps^2 - 74i\omega \eps^2 + 7\eps^4=0~.
%\label{eq:}
%
\end{align}
The pole $\calQ$ is an $O(\omega^2)$ expression, so there are two hydrodynamic poles instead of three for the \HSF. %They correspond to $\mfA_x$ and $\dpsi_+$, and $\mfA_t$ does not have a hydrodynamic pole.

One can obtain $\calQ=0$ poles approximately. 
When $\eps \ll q$, 
\begin{subequations}
%\label{eq:eps_small}
\begin{align}
\omega &= \frac{3-i}{10}q^2+\frac{21-37i}{120}\epsilon^2 +\cdots~,\\
\omega &= \frac{-3-i}{10}q^2+\frac{-21-37i}{120}\epsilon^2 +\cdots~,
%\label{eq:}
%
\end{align}
\end{subequations}
Similarly, when $\eps \gg q $, 
\begin{subequations}
\label{eq:q_small_HSC}
\begin{align}
\omega &= -\frac{i}{2}\eps^2+\frac{4i}{23}q^2+\cdots~,
\label{eq:HSC_scalar} \\
\omega &=-\frac{7i}{60}\eps^2-\frac{43i}{115}q^2+\cdots.~
 %\label{eq:}
%
\end{align}
\end{subequations}
For the \HSF,  there are 2 damped oscillations. As the system is away from the critical point or $\eps$ becomes larger, they become second sound poles. For the \HSC, there are 2 damped oscillations again. However, as the system is away from the critical point, they become dissipative poles. 
%The former is identified as $\mfA_x$. 
%\margin{?}
%This is because $\mfA_x$ should agree with $\calA_y$ below when $q=0$.

We already show that the \HSF\ result agrees with model F equations, so imposing the Maxwell equation to model F equations reproduces the bulk results. The generic dispersion relations by model F parameters are rather complicated, but when $\eps\gg q$,
\begin{subequations}
%\label{eq:}
\begin{align}
\omega &= - \frac{2ic_K}{\sigma_n}\eps^2+\cdots~,\\
\omega &=-i\frac{ (\tau_0+\tau_0^*)(a+\gamma^2\chi_T \eps^2)}{\tau_0\tau_0^*}+\cdots~.
%\label{eq:}
%
\end{align}
\end{subequations}
This of course reproduces the bulk result \eqref{eq:q_small_HSC}.

%%------------------
\subsection{Maxwell vector perturbation}%\label{sec:}
%%------------------

For the vector mode $\calA_y$, 
\begin{align}
\bra \delta \calJ^y \ket
&=\left(i\omega+ q^2\ln r_0 - \half \epsilon^2+\cdots \right)\calA_y~, 
% old
%\quad O(l^2)
%\label{eq:}
%
\end{align}
and the holographic semiclassical equation is
\begin{align}
\left(-\epse(k)\omega^2+\frac{1}{\mu_m(k)} q^2 \right) \calA_y &= -\half \eps^2\calA_y.
%\label{eq:}
%
\end{align}

However, for usual condensed-matter materials, $\mu_m$ is almost constant, 
%\margin{E OK}
so it may be familiar for the readers to isolate the constant part of $\mu_m(k)$, $\mu_{m0}$. In this case, the semiclassical equation gives
\begin{align}
\omega= -\frac{i}{\mu_{m0}}q^2- \frac{i}{2}\eps^2~.
%\frac{1}{\mu_{m0}}q^2 \calA_y=i\omega\calA_y - \frac{\del f}{\del \calA_y}~.
\label{eq:vector}
\end{align}
%This is the dual GL equation for the vector mode.

In the high-temperature phase or for a dynamical photon in the $\mathcal{N}=4$ SYM, $\epsilon=0$, so
\begin{subequations}
% corrected 2026/5/10
%\label{eq:}
\begin{align}
&q^2 = i\mu_{m0}\omega~, \\
\to &q = \sqrt{\mu_{m0}\omega}(1+i) = \frac{1+i}{\delta}~, \quad 
\delta= \sqrt{\frac{2}{\mu_{m0}\omega}}~, \\
\to &E^y = E e^{-x/\delta}e^{-i\omega t+ix/\delta}~,
%\label{eq:}
%
\end{align}
\end{subequations}
where $\delta$ is called the ``skin depth." 
It represents the distance that electromagnetic field penetrates into a conductor from the vacuum.
 %(Landau, Sect.60).  ($\sqrt{2i}=1+i$.) 
 Of course, we do not have a ``surface" of the conductor in our holographic system.
%In the Neumann BC, $\mu_m \to -1/\ln r_0$.

%When $\omega,q \neq0$, the semiclassical equation gives a pole at
%\begin{align}
%
%\omega= -\frac{i}{\mu_{m0}}q^2- \frac{i}{2}\eps^2~.
%\label{eq:}
%
%\end{align}
In the low-temperature phase, when $\omega=0$, \eq{vector} becomes
\begin{align}
q^2= -\frac{\mu_{m0}}{2}\eps^2~.
%\label{eq:}
%
\end{align}
This is the Meissner effect with the penetration length $\lambda_\text{London}^2=2/(\mu_{m0} \eps^2)$. 
%When $q=0$, $\omega=-i\eps^2/2$ that agrees with \eq{HSC_scalar}.

The dual GL equation for the vector mode can be written as
\begin{subequations}
%\label{eq:}
\begin{align}
\lambda\del_t \calA_y &=  -\frac{\delta F}{\delta \calA_y}~,\\
\to -i\omega\lambda \calA_y &= -\frac{1}{\mu_{m0}}q^2 \calA_y - \frac{\del f}{\del \calA_y}~,
%\label{eq:}
%
\end{align}
\end{subequations}
when one imposes the Maxwell equation. Just like the TDGL equation or the model F equation, the left-hand-side represents a dissipative effect.
%This is the dual GL equation for the vector mode.

%%%%%%%%%
\section{Discussion}\label{sec:discussion}
%%%%%%%%%

\begin{enumerate}
\item
As we saw, we impose the holographic semiclassical equation as the boundary condition for the \HSC, but  it reduces to the Neumann boundary condition for the scalar mode at the leading order in our expansion. Then, the result of the \HSF\ implicitly contains the result of the \HSC. One would say that the denominator $\calP$ gives poles as the \HSF\ whereas the numerator $\calQ$ gives poles as the \HSC. 

\item
In this paper, we focus on a minimal \HSFHSC, but there exist other analytic solutions \cite{Herzog:2010vz,Natsuume:2018yrg,2nd,Herzog:2009ci}, and it would be interesting to carry out a similar analysis for those solutions.
\item
For example, one can consider a class of ``nonmimimal" \HSFHSCs:
\begin{subequations}
\label{eq:nonminimal}
\begin{align}
S_\text{m} &= -\frac{1}{g^2} \int d^5x \sqrt{-g} \biggl\{ \frac{1}{4}F_{MN}^2 + K|D_M\Psi|^2+V \biggr\}~,\\
K &= 1+A|\Psi|^2~,\quad
V= m^2 |\Psi|^2+B |\Psi|^4~.
%\label{eq:}
%
\end{align}
\end{subequations}
Here, $A$ and $B$ are bulk parameters. The original \HSFHSC\ is the $A=B=0$ case which we call the ``minimal" \HSFHSC. A similar analysis can be done for nonminimal \HSFHSCs, and one can show that the results agree with bulk results with the same set of model F parameters \eqref{eq:F_parameters}. 
The only difference is that one should use
\begin{align}
\epsmu=\frac{1+4A+4B}{24}\eps^2~.
%\label{eq:}
%
\end{align}
In this sense, nonminimal \HSFHSCs\ have the identical critical dynamics with the minimal one near the critical point.

\end{enumerate}

%%%%%%%%%
%\section*{Acknowledgments}%\label{sec:} % for PTEP,PRD
%%%%%%%%%

%%%%%%%%%
\acknowledgments % for JHEP
%%%%%%%%%

%\begin{acknowledgments}
I would like to thank Takashi Okamura for collaboration on the early stages of this work and for his continuous suggestions and interest throughout the work.
This research was supported in part by a Grant-in-Aid for Scientific Research (25K07291) from the Ministry of Education, Culture, Sports, Science and Technology, Japan. 
%\end{acknowledgments}

%%%%%%%%%

\appendix

%%%%%%%%%
\section{Supplementary information}%\label{sec:}
%%%%%%%%%

%%------------------
\subsection{The conventional model F equations}\label{sec:conventional}
%%------------------

In our conventions, the conventional model F equations are given by
\begin{subequations}
\label{eq:original_F}
\begin{align}
\del_t\psi &= - \tau_0^{-1}\frac{\delta F}{\delta \psi^*}  -ig_0 \psi\frac{\delta F}{\delta \rho}~, \\
\del_t \rho &= \lambda \del_i^2 \left( \frac{\delta F}{\delta \rho} \right) +2g_0 \Im \left( \psi^*\frac{\delta F}{\delta \psi^*} \right)~.
%\label{eq:}
%
\end{align}
\end{subequations}
%Here, $D_\mu=\del_\mu-ig \calA_\mu$ and 
The GL free energy is given by
\begin{align}
f &= c_K|D_i\psi|^2 - a |\psi|^2 + \frac{b'}{2}|\psi|^4+
\frac{1}{2C}\rho^2 +\gamma \rho |\psi|^2-\rho \At -(\psi J^*+\psi^* J)~. 
\end{align}

%%------------------
\subsection{Stationary limit of model F equations}\label{sec:F_to_GL}
%%------------------

Consider the $\omega=0$ limit of model F equations. In this case, one can consistently set $\dpsi^*=\dpsi, \Jc=J,\delta\calA_x=0$. Then, 
\begin{align}
\delta\rho=C(\delta\calA_t-2\gamma\eps\dpsi)~.
%\label{eq:}
%
\end{align}
Substituting the solution and the background solution $\rho_0=-C\gamma\eps^2$, the $\dpsi$-equation in the low-temperature phase becomes
\begin{align}
j= \left\{ c_Kq^2-a+3(b'-C\gamma^2)\eps^2 \right\} \dpsi+O(\delta\calA_t)~.
%\label{eq:}
%
\end{align}
If one focuses only on the order parameter, one can set
\begin{align}
b=b'-C\gamma^2=\frac{1}{48}
%\label{eq:}
%
\end{align}
with free energy
\begin{align}
f &= c_K|D_i\psi|^2 - a |\psi|^2 + \frac{b}{2}|\psi|^4 -(\psi J^*+\psi^* J)~. 
%\label{eq:}
%
\end{align}

%%------------------
\section{Static universality class of the $\mathcal{N}=4$ SYM}\label{sec:N=4}
%%------------------

This is somewhat beyond the scope of this paper, but we include it for recent renewed interests \cite{Gladden:2024ssb,Gladden:2025glw}.
For a ferromagnet, the static exponents are traditionally defined by
\begin{subequations}
%\label{eq:}
\begin{align}
&\text{specific heat: } & C_H & \propto t^{-\alpha}~, & \\
&\text{spontaneous magnetization: } & m & \propto t^{\beta}~, & (T< T_c), \\
&\text{magnetic susceptibility: } & \chi_T & \propto t^{-\gamma}~, & \\ 
&\text{critical isotherm: } & m & \propto |H|^{1/\delta}~, & (T= \Tc),\\
&\text{static susceptibility:  } & \chi(\bmr\,) & \propto e^{-r/\xi}~, & (T \neq  T_c), \\
%\\
%&                    && \propto r^{-\ds+2-\eta}~, & (T= \Tc), \\
&\text{correlation length: } & \xi & \propto t^{-\nu}~, &
\end{align}
\end{subequations}
For the standard $|\psi|^4$ mean-field theory, 
\begin{align}
(\alpha, \beta, \gamma, \delta, \nu, \eta) = \left(0, \frac{1}{2}, 1, 3,  \frac{1}{2}, 0\right)~.
%\label{eq:}
%
\end{align}

From the 10-dimensional point of view, the $\mathcal{N}=4$ SYM is dual to SAdS$_5 \times S^5$, and one can add angular momenta along $S^5$. $S^5$ has the $SO(6)$ symmetry which is rank 3, so one can add 3 independent angular momenta. From the 5-dimensional bulk point of view, these angular momenta become Kaluza-Klein charges. They correspond to R-charges of the $\mathcal{N}=4$ SYM.

The $\mathcal{N}=4$ SYM undergoes a second-order phase transition at a finite chemical, and a charge shows a singular behavior at the critical point, so the charge corresponds to the order parameter.
As computed in Refs.~\cite{Gubser:1998jb,Cai:1998ji,Cvetic:1999rb,Maeda:2008hn}, the theory has static exponents 
\begin{align}
(\alpha,\beta,\gamma,\delta,\nu,\eta) = \biggl(-\half,\half,\half,2,\frac{1}{4},0 \biggr)~.
\label{eq:N=4_exponents}
\end{align}
These exponents are less familiar, but this is a $m^3$ mean-field theory with a slight change:
%\footnote{This possibility was pointed out by Masafumi Fukuma and Hikaru Kawai. We thank them.}
%The $m^3$ mean-field theory is given by %the continuum limit of the Potts model which describes percolation:
\begin{align}
f= \frac{1}{2} (\del_i m)^2-\frac{1}{2}am^2+\frac{1}{3}m^3 +\cdots -mH~,
%a = a_0(t-1)+\cdots~,
%\label{eq:}
%
\end{align}
where $t$ is the normalized temperature $t:=T/T_c$. 

Usually, one would take $ a = a_0(t-1)+\cdots$. Then, the theory describes the continuum limit of the Potts model which describes percolation:
\begin{align}
(\alpha,\beta,\gamma,\delta,\nu,\eta) = \biggl(-1,1,1,2,\frac{1}{2},0 \biggr)~.
%\label{eq:}
%
\end{align}
In gauge theories, a deformed $\mathcal{N}=4$ SYM belongs to this universality class \cite{Banks:2015aca} as we pointed out to the authors previously.

For the $\mathcal{N}=4$ SYM, take $a = a_0(t-1)^{1/2}+\cdots$. It is less familiar, but it gives the correct critical exponents:
\begin{enumerate}
\item
First, consider the homogeneous case where $\del_i m=0$. 
The saddle point is given by $\delta f=m(-a+bm)=0$, so $m=a/b$. 
This gives $\beta=1/2$.

\item
Then, the on-shell free energy $\underline{f}$ is given by $\underline{f} \propto -a^3/b^2.$ The specific heat is $C_H = -T\del_T^2 \underline{f} \propto (t-1)^{1/2}$.  
This gives $\alpha=-1/2$.

\item 
When one adds the source term $H$, the equation of motion is given by $\delta f=-am+bm^2-H=0$, so $m^2\propto H$ at the saddle point. 
This gives $\delta=1/2$.

\item
Differentiating the equation with respect to $H$ gives the susceptibility $\chi_T$:
\begin{align}
\del_H \delta f=(-a+2bm)\del_Hm-1=0~\to \chi_T =\del_H m \propto a^{-1}~.
%\label{eq:}
%
\end{align}
This gives $\gamma=1/2$.

\item
In the stationary case, the equation of motion is given by $0=\del^2m-am$, where we consider the high-temperature phase. In the momentum space $m\propto e^{iqx}$,
\begin{align}
0=(q^2-a)m \to \xi_>^2 = -1/q^2 =1/a~.
%\label{eq:}
%
\end{align}
This gives $\nu=1/2$.

\end{enumerate}

%Now, this is not conventional, but let us take $a = a_0(t-1)^{1/2}+\cdots$. Then, one obtains static exponents \eqref{eq:N=4_exponents}.

However, the GL theory also implies that the system is unstable at the critical point and undergoes the first-order phase transition. The bulk solution at the low-temperature phase is unknown as far as we are aware. 

Also, the $m^3$ potential is unbounded below. But this cannot be judged from the GL theory alone. The GL theory is an effective theory just like the \HSC, and there should be an infinite number of higher order terms $m^n$. Presumably, those higher order terms may fix the problem, and the complete potential may be bounded below.  

% v2
%%%%%%%%%
\section{Comparison with Ref.~\cite{Bu:2026mxr} (added in v2)}\label{sec:comparison}
%%%%%%%%%

Ref.~\cite{Bu:2026mxr} studies the same \HSF/\HSC\ as our systems, so we compare our results with theirs (based on their v1).% 
\footnote{For conventions, they set the horizon radius $r_h=1$, so they work in the unit $\pi T=1$ which is the same as ours. For the holographic renormalization, they choose the renormalization scale $\ln r_h=1$ from their Eq.~(5.4). Their AdS/CFT dictionary for $\Psi$ (5.6) is the same as ours up to an overall sign, which does not affect analysis. }
\begin{itemize}

\item
The paper uses the Schwinger-Keldysh formalism and its holographic implementation developed in 
Ref.~\cite{Crossley:2015evo}.

\item
The paper carries out the holographic computation in the high-temperature phase [see, \eg, their Eq.~(5.10)] and does not carries out the computation in the low-temperature phase. The complete time-dependent computation in the low-temperature phase is much more involved.

\item
We impose the ``holographic semiclassical equation" 
\eqref{eq:semiclassical} 
for the \HSC, but it reduces to the Neumann boundary condition at the leading order in the expansion. They impose the Neumann boundary condition from the beginning.

\item
We use the $(\omega,q^2)$-expansions that respects model F symmetry whereas they use the $(\omega,q)$-expansion. This may be a reason why our dispersion relations are different from theirs [their Eq.~(4.7)]. 

\item
Their Hamiltonian (free energy) (A.10) and model F parameters (A.11) do not reproduce our bulk results  \eqref{eq:solution_mfAt}. This may come from the fact that they set all external sources to zero to obtain model F parameters.

\end{itemize}

\footnotesize

%\newpage


\begin{thebibliography}{}

%%% AdS/CFT %%%

%\cite{Maldacena:1997re}
\bibitem{Maldacena:1997re}
  J.~M.~Maldacena,
  ``The Large N limit of superconformal field theories and supergravity,''
  Int.\ J.\ Theor.\ Phys.\  {\bf 38} (1999) 1113
   [Adv.\ Theor.\ Math.\ Phys.\  {\bf 2} (1998) 231]
%  doi:10.1023/A:1026654312961
  [hep-th/9711200].
  %%CITATION = doi:10.1023/A:1026654312961;%%

%\cite{Witten:1998qj}
\bibitem{Witten:1998qj}
  E.~Witten,
  ``Anti-de Sitter space and holography,''
  Adv.\ Theor.\ Math.\ Phys.\  {\bf 2} (1998) 253
  [hep-th/9802150].
  %%CITATION = HEP-TH/9802150;%%

%\cite{Witten:1998zw}
\bibitem{Witten:1998zw}
  E.~Witten,
  ``Anti-de Sitter space, thermal phase transition, and confinement in gauge theories,''
  Adv.\ Theor.\ Math.\ Phys.\  {\bf 2} (1998) 505
  [hep-th/9803131].
  %%CITATION = HEP-TH/9803131;%%
  
%\cite{Gubser:1998bc}
\bibitem{Gubser:1998bc}
  S.~S.~Gubser, I.~R.~Klebanov and A.~M.~Polyakov,
  ``Gauge theory correlators from noncritical string theory,''
  Phys.\ Lett.\ B {\bf 428} (1998) 105
%  doi:10.1016/S0370-2693(98)00377-3
  [hep-th/9802109].
  %%CITATION = doi:10.1016/S0370-2693(98)00377-3;%%

%%% holography textbooks %%%

%\cite{CasalderreySolana:2011us}
\bibitem{CasalderreySolana:2011us}
  J.~Casalderrey-Solana, H.~Liu, D.~Mateos, K.~Rajagopal and U.~A.~Wiedemann,
  \textit{Gauge/String Duality, Hot QCD and Heavy Ion Collisions} (Cambridge Univ.\ Press, 2014)
  [arXiv:1101.0618 [hep-th]].
  %%CITATION = ARXIV:1101.0618;%%

%\cite{Natsuume:2014sfa}
\bibitem{Natsuume:2014sfa}
  M.~Natsuume,
  \textit{AdS/CFT Duality User Guide},
  Lecture Notes in Physics Vol. 903 (Springer Japan, Tokyo, 2015) 
%  Lect.\ Notes Phys.\  {\bf 903} (2015) 
%  doi:10.1007/978-4-431-55441-7
  [arXiv:1409.3575 [hep-th]].
  %%CITATION = doi:10.1007/978-4-431-55441-7;%%

%\cite{Ammon:2015wua}
\bibitem{Ammon:2015wua}
  M.~Ammon and J.~Erdmenger,
  \textit{Gauge/gravity duality : Foundations and applications}
  (Cambridge Univ.\ Press, 2015).
  %%CITATION = INSPIRE-1376202;%%

%\cite{Zaanen:2015oix}
\bibitem{Zaanen:2015oix}
  J.~Zaanen, Y.~W.~Sun, Y.~Liu and K.~Schalm,
  \textit{Holographic Duality in Condensed Matter Physics}
  (Cambridge Univ.\ Press, 2015).
  %%CITATION = INSPIRE-1384852;%%

%\cite{Hartnoll:2016apf}
\bibitem{Hartnoll:2016apf}
  S.~A.~Hartnoll, A.~Lucas and S.~Sachdev,
  \textit{Holographic quantum matter}
  (The MIT Press, 2018) 
  [arXiv:1612.07324 [hep-th]].
  %%CITATION = ARXIV:1612.07324;%%

%\cite{Baggioli:2019rrs}
\bibitem{Baggioli:2019rrs}
M.~Baggioli,
\textit{Applied Holography: A Practical Mini-Course},
SpringerBriefs in Physics (Springer, 2019)
%doi:10.1007/978-3-030-35184-7
[arXiv:1908.02667 [hep-th]].

%%% HSC %%%

%\cite{Gubser:2008px}
\bibitem{Gubser:2008px}
S.~S.~Gubser,
``Breaking an Abelian gauge symmetry near a black hole horizon,''
Phys. Rev. D \textbf{78} (2008), 065034
%doi:10.1103/PhysRevD.78.065034
[arXiv:0801.2977 [hep-th]].

%\cite{Hartnoll:2008vx}
\bibitem{Hartnoll:2008vx}
S.~A.~Hartnoll, C.~P.~Herzog and G.~T.~Horowitz,
``Building a Holographic Superconductor,''
Phys. Rev. Lett. \textbf{101} (2008), 031601
%doi:10.1103/PhysRevLett.101.031601
[arXiv:0803.3295 [hep-th]].

%\cite{Hartnoll:2008kx}
\bibitem{Hartnoll:2008kx}
S.~A.~Hartnoll, C.~P.~Herzog and G.~T.~Horowitz,
``Holographic Superconductors,''
JHEP \textbf{12} (2008), 015
%doi:10.1088/1126-6708/2008/12/015
[arXiv:0810.1563 [hep-th]].

%%%

%\bibitem{tinkham}
%M.~Tinkham, \textit{Introduction to superconductivity: second edition} (Dover Publications, 2004).

%\cite{Herzog:2010vz}
\bibitem{Herzog:2010vz}
  C.~P.~Herzog,
  ``An Analytic Holographic Superconductor,''
  Phys.\ Rev.\ D {\bf 81} (2010) 126009
%  doi:10.1103/PhysRevD.81.126009
  [arXiv:1003.3278 [hep-th]].
  %%CITATION = doi:10.1103/PhysRevD.81.126009;%%

%\cite{Natsuume:2018yrg}
\bibitem{Natsuume:2018yrg}
M.~Natsuume and T.~Okamura,
``Holographic Lifshitz superconductors: Analytic solution,''
Phys. Rev. D \textbf{97} (2018) no.6, 066016
%doi:10.1103/PhysRevD.97.066016
[arXiv:1801.03154 [hep-th]].

%\cite{Natsuume:2022kic}
\bibitem{Natsuume:2022kic}
M.~Natsuume and T.~Okamura,
``Holographic Meissner effect,''
Phys. Rev. D \textbf{106} (2022) no.8, 086005
%doi:10.1103/PhysRevD.106.086005
[arXiv:2207.07182 [hep-th]].

%\cite{Natsuume:2024ril}
\bibitem{first}
M.~Natsuume,
``What is the Dual Ginzburg-Landau Theory for Holographic Superconductors?,''
PTEP \textbf{2025} (2025) no.2, 023B08
%doi:10.1093/ptep/ptaf018
[arXiv:2407.13956 [hep-th]].

%\cite{Natsuume:2024sic}
\bibitem{2nd}
M.~Natsuume,
``The dual Ginzburg-Landau theory for a holographic superconductor: finite coupling corrections,''
JHEP \textbf{11} (2024), 107
%doi:10.1007/JHEP11(2024)107
[arXiv:2409.18323 [hep-th]].

%\cite{Natsuume:2026jgm}
%\bibitem{Natsuume:2026jgm}
\bibitem{2nd2}
M.~Natsuume,
``Comments on higher-derivative corrections in the AdS/CFT duality,''
[arXiv:2605.15478 [hep-th]].

%%% holographic F %%%
%\cite{Flory:2022uzp}
\bibitem{Flory:2022uzp}
M.~Flory, S.~Grieninger and S.~Morales-Tejera,
``Critical and near-critical relaxation of holographic superfluids,''
Phys. Rev. D \textbf{110} (2024) no.2, 026019
%doi:10.1103/PhysRevD.110.026019
[arXiv:2209.09251 [hep-th]].

%\cite{Donos:2022qao}
\bibitem{Donos:2022qao}
A.~Donos and P.~Kailidis,
``Nearly critical holographic superfluids,''
JHEP \textbf{12} (2022), 028
[erratum: JHEP \textbf{07} (2023), 232]
%doi:10.1007/JHEP12(2022)028
[arXiv:2210.06513 [hep-th]].

%\cite{Donos:2025jxb}
\bibitem{Donos:2025jxb}
A.~Donos and P.~Kailidis,
``Critical dynamics of superfluids,''
JHEP \textbf{02} (2026), 121
%doi:10.1007/JHEP02(2026)121
[arXiv:2510.20750 [hep-th]].

%%% 
% v2
\bibitem{cardy}
J.~Cardy,
\textit{Scaling and renormalization in statistical physics}
(Cambridge Univ. Press, 1996).

% v12
%\cite{Compere:2008us}
\bibitem{Compere:2008us}
G.~Compere and D.~Marolf,
``Setting the boundary free in AdS/CFT,''
Class. Quant. Grav. \textbf{25} (2008), 195014
%doi:10.1088/0264-9381/25/19/195014
[arXiv:0805.1902 [hep-th]].

% v12
%\cite{Domenech:2010nf}
\bibitem{Domenech:2010nf}
O.~Domenech, M.~Montull, A.~Pomarol, A.~Salvio and P.~J.~Silva,
``Emergent Gauge Fields in Holographic Superconductors,''
JHEP \textbf{08} (2010), 033
%doi:10.1007/JHEP08(2010)033
[arXiv:1005.1776 [hep-th]].

%\cite{Breitenlohner:1982bm}
%\bibitem{Breitenlohner:1982bm}
%  P.~Breitenlohner and D.~Z.~Freedman,
%  ``Positive Energy in anti-De Sitter Backgrounds and Gauged Extended Supergravity,''
%  Phys.\ Lett.\  {\bf 115B} (1982) 197.
%  doi:10.1016/0370-2693(82)90643-8
  %%CITATION = doi:10.1016/0370-2693(82)90643-8;%%
  
%%% universality class %%%

%\cite{Maeda:2009wv}
\bibitem{Maeda:2009wv}
K.~Maeda, M.~Natsuume and T.~Okamura,
``Universality class of holographic superconductors,''
Phys. Rev. D \textbf{79} (2009), 126004
%doi:10.1103/PhysRevD.79.126004
[arXiv:0904.1914 [hep-th]].

\bibitem{hohenberg_halperin}
  P.C. Hohenberg and B.I. Halperin,
``Theory of dynamic critical phenomena,"
  Rev.\ Mod.\ Phys.\  {\bf 49} (1977) 435.
 
 %\cite{Maeda:2008hn}
\bibitem{Maeda:2008hn}
K.~Maeda, M.~Natsuume and T.~Okamura,
``Dynamic critical phenomena in the AdS/CFT duality,''
Phys. Rev. D \textbf{78} (2008), 106007
%doi:10.1103/PhysRevD.78.106007
[arXiv:0809.4074 [hep-th]].


%\cite{Buchel:2010gd}
\bibitem{Buchel:2010gd}
A.~Buchel,
``Critical phenomena in N=4 SYM plasma,''
Nucl. Phys. B \textbf{841} (2010), 59-99
%doi:10.1016/j.nuclphysb.2010.07.017
[arXiv:1005.0819 [hep-th]].

%\cite{Rajagopal:1992qz}
\bibitem{Rajagopal:1992qz}
K.~Rajagopal and F.~Wilczek,
``Static and dynamic critical phenomena at a second order QCD phase transition,''
Nucl. Phys. B \textbf{399} (1993), 395-425
%doi:10.1016/0550-3213(93)90502-G
[arXiv:hep-ph/9210253 [hep-ph]].

%\cite{Son:2004iv}
\bibitem{Son:2004iv}
D.~T.~Son and M.~A.~Stephanov,
``Dynamic universality class of the QCD critical point,''
Phys. Rev. D \textbf{70} (2004), 056001
%doi:10.1103/PhysRevD.70.056001
[arXiv:hep-ph/0401052 [hep-ph]].

 %\cite{Natsuume:2010bs}
\bibitem{Natsuume:2010bs}
M.~Natsuume and T.~Okamura,
``Dynamic universality class of large-N gauge theories,''
Phys. Rev. D \textbf{83} (2011), 046008
%doi:10.1103/PhysRevD.83.046008
[arXiv:1012.0575 [hep-th]].
 
%%%

% v2
%\cite{Fuini:2015hba}
\bibitem{Fuini:2015hba}
J.~F.~Fuini and L.~G.~Yaffe,
``Far-from-equilibrium dynamics of a strongly coupled non-Abelian plasma with non-zero charge density or external magnetic field,''
JHEP \textbf{07} (2015), 116
%doi:10.1007/JHEP07(2015)116
[arXiv:1503.07148 [hep-th]].

%\cite{Amado:2009ts}
\bibitem{Amado:2009ts}
I.~Amado, M.~Kaminski and K.~Landsteiner,
``Hydrodynamics of Holographic Superconductors,''
JHEP \textbf{05} (2009), 021
%doi:10.1088/1126-6708/2009/05/021
[arXiv:0903.2209 [hep-th]].

%\cite{Policastro:2002se}
\bibitem{Policastro:2002se}
G.~Policastro, D.~T.~Son and A.~O.~Starinets,
``From AdS / CFT correspondence to hydrodynamics,''
JHEP \textbf{09} (2002), 043
%doi:10.1088/1126-6708/2002/09/043
[arXiv:hep-th/0205052 [hep-th]].

%\cite{Herzog:2009ci}
\bibitem{Herzog:2009ci}
C.~P.~Herzog and S.~S.~Pufu,
``The Second Sound of SU(2),''
JHEP \textbf{04} (2009), 126
%doi:10.1088/1126-6708/2009/04/126
[arXiv:0902.0409 [hep-th]].

%%% N=4
%\cite{Gubser:1998jb}
\bibitem{Gubser:1998jb}
  S.~S.~Gubser,
  ``Thermodynamics of spinning D3-branes,''
  Nucl.\ Phys.\  B {\bf 551} (1999) 667
  [arXiv:hep-th/9810225].
  %%CITATION = NUPHA,B551,667;%%

%\cite{Cai:1998ji}
\bibitem{Cai:1998ji}
  R.~G.~Cai and K.~S.~Soh,
  ``Critical behavior in the rotating D-branes,''
  Mod.\ Phys.\ Lett.\  A {\bf 14} (1999) 1895
  [arXiv:hep-th/9812121].
  %%CITATION = MPLAE,A14,1895;%%

%\cite{Cvetic:1999rb}
\bibitem{Cvetic:1999rb}
  M.~Cvetic and S.~S.~Gubser,
  ``Thermodynamic stability and phases of general spinning branes,''
  JHEP {\bf 9907} (1999) 010
  [arXiv:hep-th/9903132].
  %%CITATION = JHEPA,9907,010;%%

%\cite{Banks:2015aca}
\bibitem{Banks:2015aca}
E.~Banks and J.~P.~Gauntlett,
``A new phase for the anisotropic N=4 super Yang-Mills plasma,''
JHEP \textbf{09} (2015), 126
%doi:10.1007/JHEP09(2015)126
[arXiv:1506.07176 [hep-th]].

%\cite{Gladden:2024ssb}
\bibitem{Gladden:2024ssb}
L.~Gladden, V.~Ivo, P.~Kovtun and A.~O.~Starinets,
``Instability in ${\cal N}=4$ supersymmetric Yang-Mills theory at finite density,''
[arXiv:2412.12353 [hep-th]].

%\cite{Gladden:2025glw}
\bibitem{Gladden:2025glw}
L.~Gladden, V.~Ivo, P.~K.~Kovtun and A.~O.~Starinets,
``Hydrodynamics with multiple charges and holography,''
JHEP \textbf{12} (2025), 173
%doi:10.1007/JHEP12(2025)173
[arXiv:2507.21346 [hep-th]].

%%%
%\cite{Bu:2026mxr}
\bibitem{Bu:2026mxr}
Y.~Bu and Z.~Yang,
``Effective Field Theory for Superconducting Phase Transitions,''
[arXiv:2604.02133 [hep-th]].

% v12
%\cite{Crossley:2015evo}
\bibitem{Crossley:2015evo}
M.~Crossley, P.~Glorioso and H.~Liu,
``Effective field theory of dissipative fluids,''
JHEP \textbf{09} (2017), 095
%doi:10.1007/JHEP09(2017)095
[arXiv:1511.03646 [hep-th]].

\end{thebibliography}
\end{document}